\documentclass[fleqn,usenatbib]{mnras}

\usepackage{newtxtext,newtxmath}
\usepackage[T1]{fontenc}
\usepackage{ae,aecompl}
\usepackage{graphicx}
\usepackage{amsmath}
\usepackage{breqn}
\usepackage[normalem]{ulem}	
\usepackage{wasysym} 
\usepackage{cleveref}
\usepackage{xcolor}

\usepackage{caption}
\usepackage{subcaption}

\crefname{Section}{\S}{\S\S}

\title[The closure radius]{Feedback reshapes the baryon distribution within haloes, in halo outskirts, and beyond: the closure radius from dwarfs to massive clusters}

\author[Ayromlou, Nelson, \& Pillepich]{Mohammadreza Ayromlou,$^{1}$\thanks{E-mail: ayromlou@uni-heidelberg.de}
Dylan Nelson,$^{1}$
Annalisa Pillepich$^{2}$
\\\\
$^{1}$Universit{\"a}t Heidelberg, Zentrum f{\"u}r Astronomie, Institut f{\"u}r theoretische Astrophysik, Albert-Ueberle-Str. 2, 69120 Heidelberg, Germany\\
$^{2}$Max-Planck-Institut f{\"u}r Astronomie, K{\"o}nigstuhl 17, 69117 Heidelberg, Germany
}

\date{}
\pubyear{2022}

\begin{document}
\label{firstpage}
\pagerange{\pageref{firstpage}--\pageref{lastpage}}
\maketitle

\begin{abstract}
We explore three sets of cosmological hydrodynamical simulations, IllustrisTNG (TNG50, TNG100, TNG300), EAGLE, and SIMBA, to investigate the physical processes impacting the distribution of baryons in and around haloes across an unprecedented mass range of $10^8 < M_{\rm 200c}/{\rm M_{\odot}} < 10^{15}$, from the halo centre out to scales as large as $30\,R_{\rm 200c}$. We demonstrate that baryonic feedback mechanisms significantly redistribute gas, lowering the baryon fractions inside haloes while simultaneously accumulating this material outside the virial radius. To understand this large-scale baryonic redistribution and identify the dominant physical processes responsible, we examine several variants of TNG that selectively exclude stellar and AGN feedback, cooling, and radiation. We find that heating from the UV background in low-mass haloes ($10^{8} \leq M_{\rm 200c}/{\rm M_{\odot}}<10^{10}$), stellar feedback in intermediate-mass haloes ($10^{10} \leq M_{\rm 200c}/{\rm M_{\odot}}<10^{12}$), and AGN feedback in groups ($10^{12} \leq M_{\rm 200c}/{\rm M_{\odot}}<10^{14}$) are the dominant processes. Galaxy clusters ($M_{\rm 200c}/{\rm M_{\odot}}>10^{14}$) are the least influenced by these processes on large scales. We introduce a new halo mass-dependent characteristic scale, the closure radius $R_{\rm c}$, within which all baryons associated with haloes are found. For groups and clusters, we introduce a universal relation between this scale and the halo baryon fraction: $R_{\rm c}/R_{\rm 200c,500c} - 1 = \beta(z) (1 - f_{\rm b}(<R_{\rm 200c,500c})/f_{\rm b,cosmic})$, where $\beta(z) = \alpha\,(1+z)^\gamma$, and $\alpha$ and $\gamma$ are free parameters fit using the simulations. Accordingly, we predict that all baryons associated with observed X-ray haloes can be found within $R_{\rm c}\sim 1.5-2.5 R_{\rm 200c}$. Our results can be used to constrain theoretical models, particularly the physics of supernova and AGN feedback, as well as their interplay with environmental processes, through comparison with current and future X-ray and SZ observations.
\end{abstract}

\begin{keywords}
galaxies: formation -- galaxies: evolution -- large-scale structure of Universe -- methods: analytical -- methods: numerical -- methods: observational
\end{keywords}


\section{Introduction}
\label{sec: introduction}

The $\rm \Lambda CDM$ model is currently the most observationally supported theory describing the expanding Universe. Observations of the cosmic microwave background infer the energy density of the Universe to consist of, at the current time, $\Omega_{\rm \Lambda}\sim 0.69$ (dark energy), $\Omega_{\rm m} \sim 0.31$ (matter), and less than one per cent from other components. The matter density of the Universe itself is comprised of $f_{\rm b,cosmic}\sim 16\%$ baryons, and $f_{\rm DM,cosmic}\sim 84\%$ dark matter \citep[][]{Spergel2003First,planck2015_xiii}.

According to the hierarchical structure formation model, baryons fall into the potential wells of dark matter haloes \citep[]{white1978core,white1991galaxy}. They then radiate energy, cool, and eventually form stars and galaxies. The dark matter halo and its baryonic content grow primarily through mergers with other haloes as well as smooth accretion from the intergalactic medium \cite[see][for a full discussion]{mo2010galaxy}. This basic scenario leads to structures within which baryons and dark matter are generally coupled, with the baryon fraction $f_{\rm b}$ close to the cosmic value. In contrast, many observations report that haloes may not contain all their associated baryons, at least within some characteristic halo boundary \citep[e.g.,][]{Persic1992baryon,Fukugita2004Cosmic}.

In particular, \cite{McGaugh2010cosmological} employed several observations \citep[][]{McGaugh2005Baryonic,Stark2009First,Trachternach2009baryonic,Walker2009Universal} to infer that haloes across a broad mass range, from Local Group dwarf satellites to low-mass clusters, have a baryon fraction within $R_{\rm 500c}$ lower than the cosmic value. More massive clusters with $M_{\rm 200c}/{\rm M_{\odot}} \gtrsim 10^{14.5}$, however, typically have a baryon fraction close to the cosmic value \cite[][]{giodini2009stellar}. Recently, X-ray surveys of groups and clusters have extensively measured the halo gas fraction within $R_{\rm 500c}$. For massive galaxy clusters, this value is close to the cosmic baryon fraction \citep[e.g. see][]{chiu2018baryon} whereas for groups it goes down to half the cosmic value \citep[e.g. see][]{lovisari2015scaling}. Consequently, these observations do not detect all the baryonic matter which should exist within the halo boundary \citep[see also][]{Shull2012Baryon,sanderson2013baryon}.

This discrepancy between the early-time observations of the baryon fraction from the CMB on large spatial scales (e.g. \citealt{komatsu2011seven,Planck2020cosmic}) and the late-time observations from the X-ray and Sunyaev Zel'dovich (SZ) surveys within halo scales \citep[e.g.,][]{gonzalez2013galaxy,Eckert2021Feedback} gives rise to one of the toughest challenges for $\rm \Lambda CDM$, often referred to as the missing baryons problem. Understanding the distribution of baryons, and the physical processes impacting them, within and especially in the outskirts of haloes, is the topic of this paper.

Cosmological simulations and observations propose that the baryons missing from haloes should reside somewhere outside a halo boundary, such as $R_{\rm 200c}$ or $R_{\rm 500c}$ \citep[e.g.,][]{Cen1999Where,Haider2016Large-scale,Ayromlou2020Comparing}. This implies an excess of baryons in the outskirts of haloes as well as within and around filamentary structures in the cosmic web, i.e. in the warm-hot gas in the intergalactic medium (WHIM, e.g., see \citealt{Penton2004Local,Kull1999Detection,Eckert2015Warm-hot}). As a result, halo outskirts and the WHIM are both powerful regimes for studying both galaxy evolution and large-scale structure cosmology \citep[see][for reviews]{Walker2019Physics,Nicastro2022Absorption}.

The current sensitivity of observational probes makes it challenging to make measurements in many regions of interest. Nevertheless, many observational efforts have been made to undertake a baryon census, primarily through X-ray and SZ surveys. For example, among others, \cite{chiu2018baryon} analyse Chandra X-ray observations of 91 SZ selected galaxy clusters and find baryon fractions lower than the cosmic value. They also infer a trend between the halo baryon fraction and the halo mass, with more massive haloes having higher baryon fractions. More recent works have extended the study of the gas even out to the outskirts of galaxy clusters \citep[e.g. see][for the Coma cluster]{Churazov2021TempestuousC}.

At larger distances, \cite{Nicastro2018Observations} claim a detection of the diffuse WHIM based on two OVII absorbers. Extrapolating their detections to all gas phases, they conclude that their results account for all baryons, although with considerable uncertainties. This was challenged by \cite{Johnson2019Physical} who argue that the signal may arise from sources other than the WHIM itself.

Furthermore, the Sunyaev-Zel'dovich effect \citep{sunyaev1972observations} due to hot gas opens another window to the baryonic content of the Universe. For example, observations made by the Atacama Cosmology Telescope \citep[ACT,][]{Hilton2021Atacama,Orlowski-Scherer2021Atacama} and South Pole Telescope \citep[SPT,][]{Carlstrom2011Meter,Bleem2015Galaxy} provide SZ maps of galaxy clusters. These probe the thermodynamic properties of the gas and constrain the large-scale distribution of baryons \citep[e.g.][]{Amodeo2021Atacama,Meinke2021Thermal}. Searching for missing baryons, \cite{deGraaff2019Probing} analysed the Planck thermal Sunyaev-Zel'dovich (tSZ) signal \citep{Planck2016tSZ} to detect the WHIM. They argued that $\sim 11 \pm 7\%$ of the total baryonic content of the Universe is embedded in filaments \citep[see also][]{Tanimura2020Density}. Additionally, \cite{lim2020detection} studied the kinetic Sunyaev-Zel'dovich (kSZ) signal using the Planck data \citep[][]{Planck2016kSZ} for haloes with $\log_{10}(M_{\rm 200c}/M_{\odot}) \gtrsim 12$ and inferred, albeit with large uncertainties, baryon fractions within $R_{\rm 200c}$ consistent with the cosmic value.

In addition to X-ray and SZ observations, Fast Radio Bursts (FRBs) can also be used to study the baryonic density of the Universe \citep[e.g.][]{McQuinn2014Locating,Munoz2018Finding}. For example, in a recent study, \cite{Macquart2020census} uses the dispersion of a sample of localized FRBs to infer, at 95\% confidence, a cosmic baryon density consistent with the cosmic baryon fraction measured from CMB data.

There is now considerable evidence for the presence of gas beyond the halo boundary. Both observations \citep[e.g.,][]{hansen2009galaxy,Wetzel2012Galaxy,Pintos-Castro2019Evolution} and simulations \citep[e.g.,][]{balogh1999differential,bahe2012competition,Ayromlou2020Comparing,Ayromlou2021Galaxy} suggest that gas stripping extends to scales well beyond the halo boundary, a signature of the presence of the gas in halo outskirts. Moreover, we have shown that the large-scale correlation between the star formation of neighbouring galaxies \citep[galactic conformity, e.g.,][]{weinmann2006properties,kauffmann2013re} arises due to environmental processes beyond the halo virial radius \citep[][]{Ayromlou2022Physical}. Overall, baryons outside of dark matter haloes are important, though observationally difficult, probes of the large-scale distribution of matter.

In contrast to observations, all $\rm \Lambda CDM$ cosmological simulations contain their full complement of baryons, by construction. In hydrodynamical simulations, the initial conditions are based on CMB observations and match the statistical properties of fluctuations in the early Universe. Given that the total amount of both baryons and dark matter are conserved in such simulations, the total baryon content remains unchanged and equal to the cosmic value -- it is only the spatial distribution of baryons that evolves. According to contemporary cosmological simulations that include feedback, the baryon fraction within the boundaries of simulated haloes, in particular, is on average lower than the cosmic value, in qualitative agreement with observations of massive groups and clusters \citep[e.g.][]{pillepich2018First, Davies2020Quenching, Ayromlou2020Comparing, Oppenheimer2021Simulating}. The dominant physical processes responsible for this are intertwined with astrophysical feedback mechanisms associated with galaxy formation and evolution.

A critical physical process in our current understanding of galaxy formation is supermassive black hole (SMBH) feedback from active galactic nuclei (AGN). Energy input from an accreting SMBH reduces its growth and quenches star formation in massive galaxies \citep[e.g.,][]{DiMatteo2005Energy,croton2006many,Shankar2009Self,terrazas20}. AGN feedback can significantly change the baryonic (and dark matter) distribution of their host haloes, either by preventing gas cooling from the hot halo and/or by ejecting gas from the inner halo to larger distances \citep[e.g., see][]{Ettori2006Baryon,Haider2016Large-scale,springel2018first,Zinger2020Ejective}.

On the other hand, star formation in galaxies with masses less than the Milky Way is instead thought to be regulated by stellar feedback \citep[e.g.,][]{Larson1974Effects,Dekel1986Origin,Stinson2006Star,Ostriker2011Maximally}.
Observational evidence suggests that stellar feedback can eject the interstellar gas into the circumgalactic medium \citep[e.g.,][]{Shapley2003Rest,Rupke2005Outflows,Weiner2009Ubiquitous,Martin2012Demographics,Rubin2014Evidence}.
These galactic-scale outflows, or winds (e.g. from supernovae explosions) are also implemented in numerical simulations of galaxy formation \citep[e.g., ][]{Oppenheimer2006Cosmological,Sales2010Feedback,Dave2013neutral,Hopkins18,Marinacci2019Simulating,nelson2019First}.

In sufficiently low-mass dark matter haloes, on the other hand, neither black hole feedback nor supernova feedback plays a role as these halos do not form galaxies.
This is due to photo-heating by the UV background field raising the temperature of diffuse gas in the environment of these haloes \citep[e.g.,][]{Efstathiou1992Suppressing,Gnedin2000Effect,Benitez-Llambay2017properties}. The pressure effects dominate the shallow gravitational potential wells, preventing the accretion of intergalactic gas \citep[e.g., see][]{Navarro1997Effects,Hambrick2011effects,PereiraWilson2022beginning}.

The interplay and collective impact of these physical processes on the large-scale distribution of baryons can be directly studied, from the theoretical point of view, with cosmological hydrodynamical simulations. This was not possible in early studies of missing baryons \citep[e.g.][]{Cen1999Where} and has only become achievable recently with modern cosmological simulations \citep[e.g.][]{vogelsberger2014Introducing,Schaye2015eagle,dave17,dubois2020} that include a roughly realistic accounting of feedback processes. 

Using the IllustrisTNG simulations in particular, \cite{Martizzi2019Baryons} study the distribution of gaseous baryons in the cosmic web. They find that at $z=0$, most of the missing baryons are in the warm-hot intergalactic medium. This is a consequence of the SMBH-driven winds implemented in TNG, which in turn trigger gas outflows that both eject gas from the innermost regions of haloes and shock heat the gas across halo scales \citep{nelson2019First, Zinger2020Ejective, truong20, Pillepich2021X-ray, Ramesh22}. 

\cite{Sorini2022How} analyse the SIMBA simulation and conclude that stellar feedback at $z>2$ and AGN feedback at $z<2$ are the dominant processes impacting the distribution of matter in massive and low-mass haloes, respectively. They also argue that the AGN feedback of SIMBA reduces the baryon fraction of groups ($\log_{10}M_{\rm 200c}/{\rm M_{\odot}}$) down to $25\%$ of the cosmic value and that the ejected baryons reside outside the halo out to $\sim 10-20 \, R_{\rm 200c}$ \citep[see also][]{Borrow2020Cosmological}.

\cite{Mitchell2022Baryonic} investigate the preventive versus ejective scenarios of the halo baryon budget in the EAGLE simulation. They find that baryons are both prevented from accreting to, and ejected from, haloes of $\log_{10}(M_{\rm 200c}/{M_{\odot}})<12$. In more massive haloes, on the other hand, most baryons are ejected beyond the halo boundary instead of being prevented from accreting onto the halo \citep[see also][]{Davies2019gas,Wright2020impact}. Moreover, \cite{Angelinelli2022Mapping} studied the distribution of baryons in the Magneticum simulation out to $10\,R_{\rm 500c} \sim 6.7 R_{\rm 200c}$ from massive haloes ($\log_{10}M_{\rm 200c}/{\rm M_{\odot}} \gtrsim 13.2$) at $z=0.25$. They report that the baryon fraction is $51\% - 87\%$ of the cosmic value in the halocentric distances of $R<R_{\rm 500c}$ and increases to $95-100\%$ at $10R_{\rm 500c}$.

In this work, we employ three different hydrodynamical simulations to investigate the distribution of baryons both within and beyond the halo boundary. By comparing IllustrisTNG, EAGLE, and SIMBA we are able to explore and marginalise over differences in current astrophysical feedback models. We introduce a new characteristic scale, the closure radius $R_{\rm c}$, within which the total baryons associated with a given dark matter halo are found. We propose a simple relation that enables the inference of $R_{\rm c}$ given observations of the halo gas fraction at much smaller scales. Furthermore, we use several variants of the TNG simulation to discover the strength and mass range of the impact of different physical processes on the distribution of baryons, as well as on our new characteristic closure scale.

This paper is structured as follows: In Section \ref{sec: Methodology}, we describe the simulations used in this study as well as the methods employed to measure the properties of matter in and around haloes. In Section \ref{sec: R_c}, we investigate the distribution of baryonic matter compared to dark matter, introduce the closure radius, and study non-gravitational physical processes that redistribute baryons in and around haloes to large scales. In Section \ref{sec: discussion}, we discuss the meaning and emergence of the closure radius and derive a fitting formula to predict its amplitude based on the observations of the halo baryon fraction. Finally, we conclude and summarise our research in Section \ref{sec: summary}.


\section{Methodology}
\label{sec: Methodology}

\subsection{Galaxy formation simulations and relevant definitions}
\label{subsec: sims}

\subsubsection{IllustrisTNG}
\label{subsubsec: TNG}

\begin{table*}
	\centering
	\caption{The properties of the simulations used in this paper. The $-1$ suffix on each TNG run denotes the highest resolution available for that volume.}
	\label{tab: data_stat}
	\begin{tabular}{|*{5}{c|}}
		\hline \hline
		\textbf{Simulation} & \textbf{Dark matter particle mass $[\rm M_{\odot}]$} & \textbf{Mean gas particle/cell mass $[\rm M_{\odot}]$} & \textbf{Box size [cMpc]} & \textbf{Code} \\
		\hline \hline
        TNG300 (-1) & $5.9\times 10^7$ & $1.1\times 10^7$ & 302.6 & AREPO  \\
		\hline
		TNG100 (-1) & $7.5\times 10^6$ & $1.4\times 10^6$ & 110.7 & AREPO \\
		\hline
		TNG50 (-1) & $4.5\times 10^5$ & $8.5\times 10^4$ & 51.7 & AREPO \\
		\hline
		TNG variations & $7.5\times 10^6$ & $1.4\times 10^6$ & 36.9 & AREPO \\
		\hline
		EAGLE & $9.7\times 10^6$ & $1.8\times 10^6$ & 100.0 & GADGET3 \\ 
		\hline
		SIMBA & $9.6\times 10^7$ & $1.8\times 10^7$ & 147.6 & GIZMO\\ 
        \hline \hline
	\end{tabular}
\end{table*}

The IllustrisTNG simulations \citep[TNG  hereafter;][]{nelson18a,pillepich2018First,springel2018first,marinacci2018first,naiman2018first}\footnote{\href{https://www.tng-project.org/}{https://www.tng-project.org/}} are a suite of large-volume cosmological galaxy formation simulations which present the next generation of the Illustris simulation \citep[][]{vogelsberger2014Introducing,genel2014introducing,sijacki2015illustris}. Based on the moving-mesh \textsc{AREPO} code \citep[][]{springel2010pur}, TNG solves the equations of gravity and magnetohydrodynamics \citep[][]{pakmor2011magnetohydrodynamics,pakmor2013simulations} and models the formation and evolution of galaxies on cosmological scales. It includes a comprehensive physical model for galaxy formation, including radiative cooling of gas, the formation of stars from the cold gas, stellar evolution, stellar feedback \citep{pillepich2018Simulating}, and physical processes relevant to supermassive black holes, including seeding, merging, and feedback \citep[][]{weinberger17}.

The TNG model has been employed to perform several simulations. Related to our work, these include three different periodic cosmological volumes with side lengths of $\sim$ 50 Mpc (TNG50, \citealt[][]{nelson2019First,pillepich19}), 100 Mpc (TNG100), and 300 Mpc (TNG300).
TNG50 (the smallest box) has the highest resolution, whereas TNG300 (the largest box) has the lowest resolution but provides the best statistics. The properties of these simulations are summarised in Table \ref{tab: data_stat}. In this work, we use the highest resolution run of each volume, combining them across the halo mass range in order obtain a statistically robust and simultaneously well resolved population. We employ TNG300 for groups and clusters ($\log_{10}M_{\rm 200c}/{M_{\odot}}\geq13$), TNG100 for haloes with ($10\leq\log_{10}M_{\rm 200c}/{M_{\odot}}<13$), and TNG50 for haloes with ($8\leq\log_{10}M_{\rm 200c}/{M_{\odot}}<10$). This is done to cover a broad range of halo mass, from dark haloes to galaxy clusters. We justify our choices in Appendix \ref{app: TNG_convergence}. Moreover, we discuss some of our results for the original Illustris simulation in Appendix \ref{app: Illustris}.

In addition to the main runs, we use several smaller `variation' simulations which change aspects of the fiducial TNG model \citep[see][]{pillepich2018Simulating}, including runs with a) no winds i.e. no stellar feedback, b) no black holes i.e. no AGN feedback, c) no stellar feedback, no black holes, and no cooling, d) non-radiative. All of these variants are performed with the TNG100 resolution in a box of $l_{\rm box} \sim 37 \,\rm Mpc$ (see Table~\ref{tab: TNG_resolution_stat}).

The calibration of the free parameters of the TNG model was carried out at the fiducial TNG100 resolution \citep[][]{pillepich2018Simulating} by comparing against several observables of the galaxy population. Namely, the star formation rate density across time and the stellar mass function, stellar-to-halo mass ratio, halo gas fraction, black hole–stellar mass relation, and the stellar sizes of galaxies at $z=0$. In all cases, the outcome of TNG was accessed by-hand, and in relative comparison with the original Illustris model -- no quantitative or e.g. algorithmic optimization was undertaken. All physical model parameters (i.e. except gravitational softening lengths and the size of the SMBH feedback region) are kept unchanged for different simulation boxes and resolutions. Consequently, simulations based on the TNG model show non-trivial numerical convergence behaviour (see \citealt[][]{pillepich2018Simulating, pillepich2018First, pillepich19} for an example of the stellar-halo mass relation). We address numerical convergence for the purposes of our analysis in Appendix \ref{app: TNG_convergence}.

\subsubsection{EAGLE}
\label{subsubsec: EAGLE}

The Evolution and Assembly of GaLaxies and their Environments model \citep[EAGLE, ][]{Schaye2015eagle,Crain2015TheEagle}\footnote{\href{http://icc.dur.ac.uk/Eagle}{http://icc.dur.ac.uk/Eagle}} uses a modified version of the \textsc{GADGET-3} smoothed particle hydrodynamics (SPH) code \citep{Springel2005Gadget} to perform galaxy formation simulations on cosmological scale boxes. In addition to solving the equations of gravity and hydrodynamics, EAGLE implements several physical processes relevant to galaxy evolution. These primarily include gas cooling, star formation, stellar feedback, and supermassive black hole-related processes such as seeding, black hole growth, and AGN feedback. The EAGLE model is calibrated for each simulation by comparing versus observations of galaxy stellar mass function, the central galaxy stellar to black hole mass relation, and galaxy sizes. We exclusively use the large 100 Mpc volume of EAGLE, and its numerical characteristics are given in Table \ref{tab: data_stat}.

\subsubsection{SIMBA}
\label{subsubsec: SIMBA}

The SIMBA hydrodynamical simulations \citep{Dave2019SIMBA}\footnote{\href{http://simba.roe.ac.uk}{http://simba.roe.ac.uk}} update the physics of the MUFASA simulation \citep{Dave2016MUFASA} to simulate galaxy formation on cosmological scales. SIMBA implements its physical processes on top of the GIZMO code \citep{hopkins15}, which employs a meshless finite mass (MFM) hydrodynamics method. Similar to TNG and EAGLE, SIMBA implements models for key relevant physical processes such as gas cooling, star formation, stellar feedback, black hole formation and growth, AGN feedback.

The SIMBA model has been used to run several simulations with different resolutions and box sizes. In this work, we exclusively use the highest resolution version of the large $\rm \sim 150 \, Mpc$ side-length volume. The dark matter and gas mass resolution of this simulation is given in Table \ref{tab: data_stat}.

\subsubsection{Halo, Subhalo, and Galaxy identification}
\label{subsubsec: FOF_SUBFIND}

For all simulations described above, we identify structures (i.e. haloes, subhaloes, and thereby galaxies) and their properties in the same manner by using versions of EAGLE and SIMBA which have been rewritten and analysed exactly as for TNG (see \citealt[][]{Nelson2019public} data release paper). This enables us to perform a uniform analysis on these simulations.\footnote{We note that due to this new style of finding the objects and the recalculation of their properties, the objects and structures of the EAGLE and SIMBA simulations may not exactly match those originally reported by these simulations, and minor differences are expected.} The Friends of Friends algorithm \citep[FOF, ][]{Davis1985TheEvolution} is applied to each simulation in order to identify groups of dark matter particles whose distances to each other are smaller than a linking length $b=0.2$. The other types of particles/cells, including gas, stars, and black holes, are are assigned to the same group as their nearest dark matter particle. Each group of particles in 3D space is called a FOF halo (or simply halo). In this work, we consider a resolution limit of $n_{\rm min}\geq 200$ particles to identify an object as a FOF halo.\footnote{Consequently, FOF haloes with fewer than 200 particles are unresolved, and we do not consider them here.} For each halo, the \textsc{SUBFIND} algorithm \citep[][]{springel2001populating} then identifies gravitationally bound substructures in 3D space, called subhaloes. By definition, each FOF halo can only have one central subhalo, and the rest of the substructures of the halo are labelled as satellite subhaloes.

Typically, there is no well-defined shape for FOF haloes. Therefore, it is not straightforward to define a boundary for the halo. Nevertheless, it is common practice, both in simulations and in observations, to consider $R_{\rm 200c}$ ($R_{\rm 500c}$), the halocentric radius within which the total matter density equals 200 (or 500) times the critical density of the Universe, as a boundary for the halo. The mass within this radius, $M_{\rm 200c}$ ($M_{\rm 500c}$), is also often taken as the halo mass.\footnote{$M_{\rm 200c}$ and $R_{\rm 200c}$ are also referred to as the virial mass and radius in the literature, although they are not exactly the same.} In this paper, we report most of our results based on these quantities. Throughout this work, whenever we need to convert $R_{\rm 200c}$ and $R_{\rm 500c}$ to each other, we follow \cite{Ayromlou2020Comparing} who used the average ratio between these quantities at fixed halo mass in TNG. When we refer to the ‘halo boundary’ we mean $R_{\rm 200c}$, and note that a halo typically extends beyond this radius in both dark matter and baryons.

\subsection{Physical processes most relevant to this study}
\label{subsec: phys_processes}

In this subsection, we briefly describe the physical processes most relevant to our study. For more details, see the papers introducing the physical prescriptions of TNG \citep{weinberger17,pillepich2018Simulating}, EAGLE \citep{Schaye2015eagle,Crain2015TheEagle}, and SIMBA \citep[][]{Dave2019SIMBA}.

\subsubsection{Black hole seeding, growth, and AGN feedback}
\label{subsubsec: AGN_feedback}

The three simulations take somewhat similar approaches for seeding supermassive black holes (SMBHs) in haloes. They run a halo finder algorithm on the fly and place a supermassive black hole with a constant mass in the halo whenever the halo mass (TNG and EAGLE) or stellar mass (SIMBA) exceeds a specific value.

In all three models, supermassive black hole growth then occurs through the accretion of matter and mergers with other supermassive black holes. In TNG, accretion is modelled using the Bondi accretion rate \citep[][]{Bondi1952spherically,Bondi1944mechanism}, which depends on SMBH mass, the local density, temperature, and the SMBH velocity relative to its local gas. EAGLE employs a modified Bondi accretion rate that accounts for the angular momentum of the gas around the SMBH \citep[see][]{Rosas-Guevara2015impact}. In SIMBA, accretion proceeds through two different channels, based on the gas temperature. Bondi accretion occurs for the warm-hot non-ISM gas ($T>10^5 \rm K$), while the cold gas accretes based on a torque-limited model, derived from the gravitational instabilities of the gaseous disk \citep[see][]{Hopkins2011analytic,Angles-Alcazar2017Gravitational}.

AGN feedback is the main physical process which quenches massive galaxies in the three models, although its implementation differs from model to model. Here we briefly describe how each simulation models AGN feedback.

\begin{itemize}
    \item TNG: Depending on the accretion rate, SMBH feedback operates in one of two modes: thermal or kinetic. These two modes do not operate simultaneously. Thermal energy from SMBH feedback is injected (continuous in time) into the gas cells in the local environment of the SMBH. This mode of AGN feedback increases the temperature of these gas cells and affects their evolution. When the accretion rate onto the SMBH is low, which is typically the case for more massive galaxies, the feedback switches to a kinetic mode, where it injects momentum and kinetic energy into surrounding gas cells.
    This energy injection, which is released in discrete events, is imparted at random orientations and is isotropic when time-averaged. However, it can result in non-isotropic large-scale outflows depending on the physical conditions of the gas in the vicinity of the SMBH \citep{weinberger17, nelson2019First, Pillepich2021X-ray}. The impact of these outflows on the distribution of gas out to large scales is one of the key results we present in this paper. Finally, in addition to these two distinct feedback modes, the TNG model also includes an AGN radiative feedback channel which operates during the radiatively efficient state (thermal mode), impacting the temperature and cooling rate of halo gas particularly at high luminosities.
    \item EAGLE: AGN feedback has only one mode, which is thermal, and implemented similarly to stellar feedback. The energy from AGN feedback is proportional to the accretion rate onto the supermassive black hole. This energy is injected stochastically by increasing the temperature of the gas particles around the SMBH by $10^{8.5}$\,K. This high temperature is designed to prevent numerical overcooling which would otherwise occur at lower temperatures.
    \item SIMBA: AGN feedback operates in one of three modes: two kinetic wind modes, referred to as ‘radiative AGN winds’ and ‘jet mode’, respectively, and one ‘X-ray’ mode. The choice of mode depends on the SMBH mass and accretion rate. In the kinetic modes, AGN feedback is implemented by probabilistically kicking individual gas particles with a high velocity which depends on the accretion rate. At high accretion rates (radiative mode feedback), the speed is $v \sim 1000 \, \rm km/s$, whereas at low accretion rates (jet mode feedback) this speed increases substantially, up to $\sim 10^4 \, \rm km/s$. These outflows are bipolar at injection, with orientation parallel to the angular momentum of the surrounding baryonic medium. When the jet mode of AGN feedback is active and the galaxy gas to stellar mass ratio is below $20\%$, an X-ray feedback mode also becomes active. Finally, at high accretion gas, kicked gas is unheated and keeps its ISM temperature, while for the jet mode kicked gas is also heated to the halo virial temperature. In both cases, AGN-driven wind particles are hydrodynamically decoupled for some time, which is not the case for either TNG or EAGLE. All energy injection is continuous in time.
\end{itemize}

\subsubsection{Stellar feedback}
\label{subsubsec: SN_feedback}

Stellar feedback is the primary mechanism for regulating star formation in low- and intermediate-mass galaxies in most galaxy formation models, including the three simulations we analyse in this work. Nevertheless, the physical model and its numerical details used to implement stellar feedback are different for each simulation. This can produce minor, to significant, differences in the integrated properties of galaxies, such as stellar masses or star formation rates. More importantly for this study, differences in stellar feedback result in substantially different distributions of baryons across multiple scales (see Section \ref{sec: R_c}). Here we briefly describe the stellar feedback model of each simulation.

\begin{itemize}
    \item TNG invokes a stochastic kinetic wind scheme, whereby collisionless wind particles carry mass, momentum, metals, and energy away from the star-forming ISM, until they recouple and return to the gas phase, based on a density or time criterion. Their direction is random, while their mass loading, velocity, and thermal energy depend on various scalings and parameters \citep[see][]{pillepich2018Simulating}. Each gas cell has an amount of energy available to produce a wind which is proportional to its instantaneous star formation rate.
    \item In the EAGLE model, stellar feedback takes place through the injection of thermal energy, which is used to increase, by a constant value, the temperature of a stochastically chosen collection of neighbouring gas particles. The injected energy is $10^{51} \, \rm erg$ for each supernova, with progenitors above $6$\,$\rm{M}_\odot$ exploding.
    \item SIMBA models stellar feedback with a similar decoupled two-phase galactic wind model as in TNG \citep[both based on][]{Springel2003}. Among ejected wind particles, $30\%$ are hot, with their temperature specified by the difference between the supernova energy and wind kinetic energy. The wind velocity and mass loading scalings are based on those from FIRE zoom-in simulations, with modifications \citep[][]{Muratov2015Gusty,Angles-Alcazar2017cosmic}.
\end{itemize}

\subsubsection{UV background radiation}
\label{subsubsec: UVB}

In all three simulations, a UV-background (UVB) radiation field is implemented to account for the interaction between the gas and photons coming from stars and quasars whose radiation is not explicitly modelled. All three simulations implement the UVB as a spatially uniform, temporally-evolving heating term. In TNG, it is incorporated at $z<6$ following the prescription of \cite[][FG11 update]{Faucher2009New}. EAGLE turns on the UVB at $z = 11.5$ based on \cite{Haardt2001Modelling} (see also \citealt[][]{Wiersma2009effect}). Finally, SIMBA uses the UVB model of \cite{Haardt2012Radiative}. TNG and SIMBA additionally account for self-shielding throughout the simulation, following \cite{Rahmati2013evolution}. Overall, photo-heating by the UVB heats diffuse gas in the environment of haloes, which slows down or stops gas accretion in low-mass haloes \citep[e.g. see][]{Efstathiou1992Suppressing,Navarro1997Effects}.

\subsection{Profile identifier method}
\label{subsec: PROFI}

Our main analysis tool are halo-centric radial profiles. For each FOF halo, we measure the radial profile of a given quantity $Q$ (e.g. mass, density, velocity) as a function of the halocentric radius $R$, from the halo centre out to $20R_{\rm 200c}$. We do this considering all cells/particles in the entire simulation volume, and do not restrict out analysis to the FOF halo alone, nor to a random subset of all particles.

We measure profiles for gas, dark matter, stars, and black holes separately. In this paper, we focus on mass and density profiles, and in a companion paper, we will extend our investigation to other important quantities such as velocities, inflow/outflow rates, metallicities, X-ray and SZ signals, temperature, and so on. Total mass profiles sum over these four components. We construct cumulative density profiles at a given halocentric distance by dividing the cumulative mass profiles by the volume of the enclosing sphere.


\begin{figure*}
    \centering
    \includegraphics[width=0.86\textwidth]{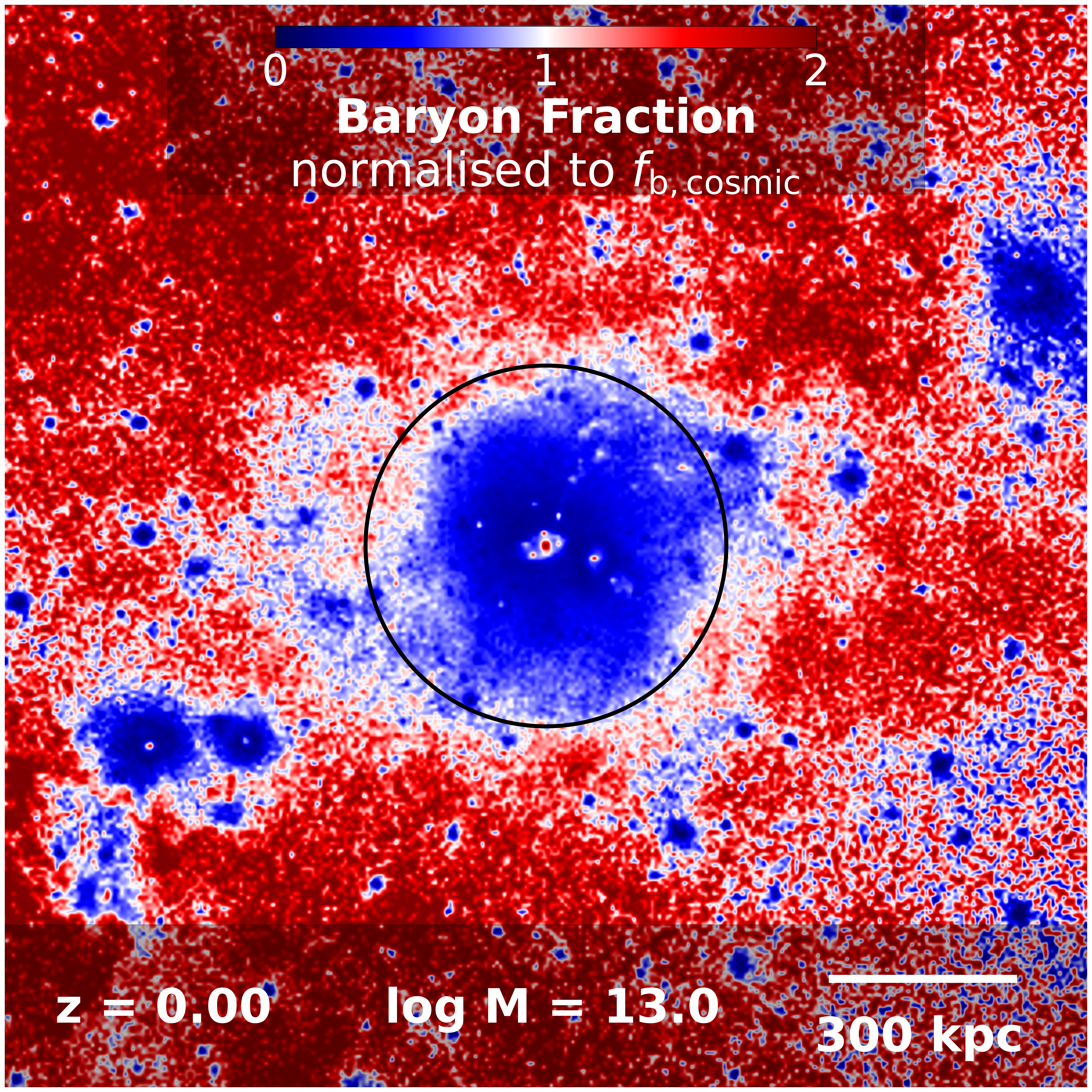}
    \includegraphics[width=0.33\textwidth]{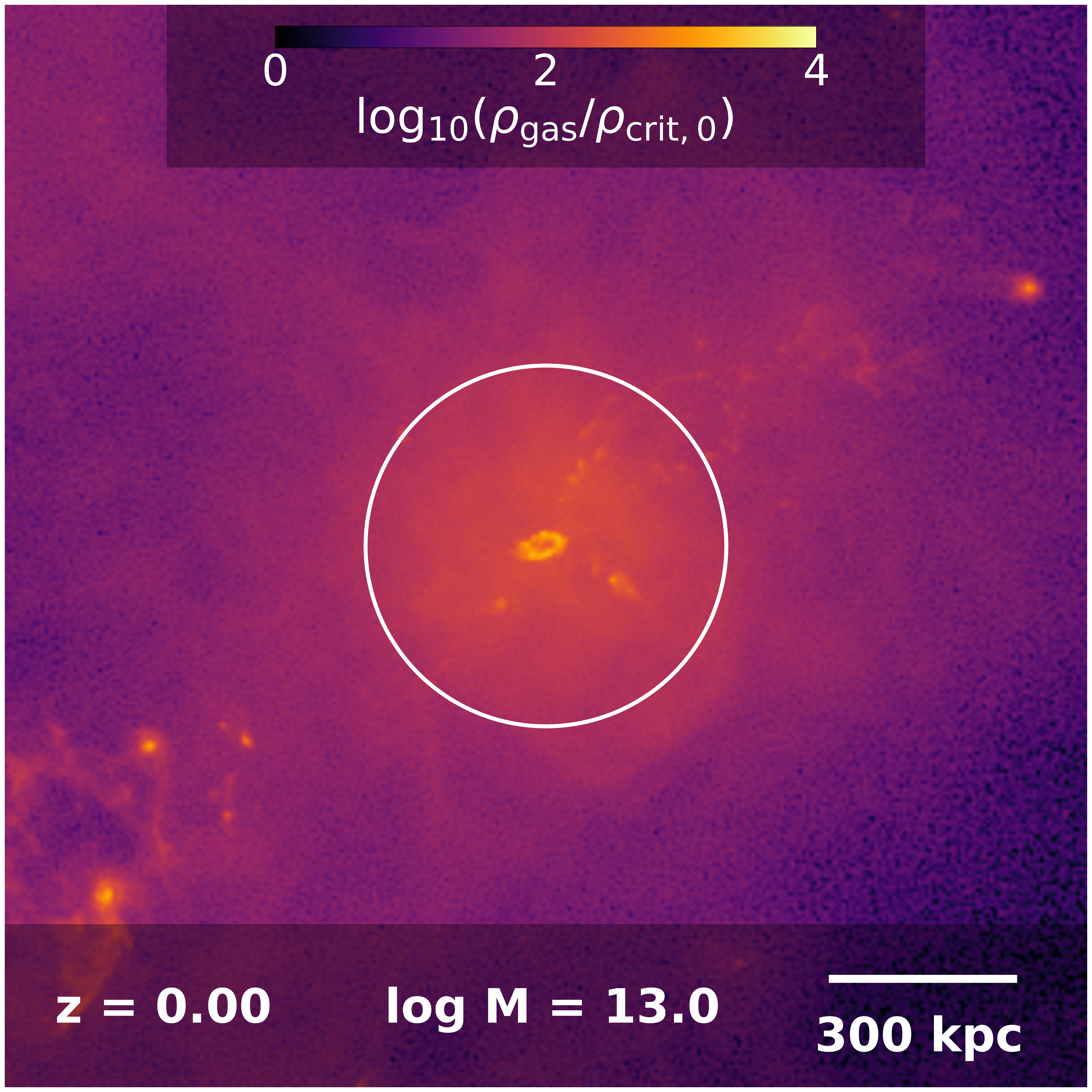}
    \includegraphics[width=0.33\textwidth]{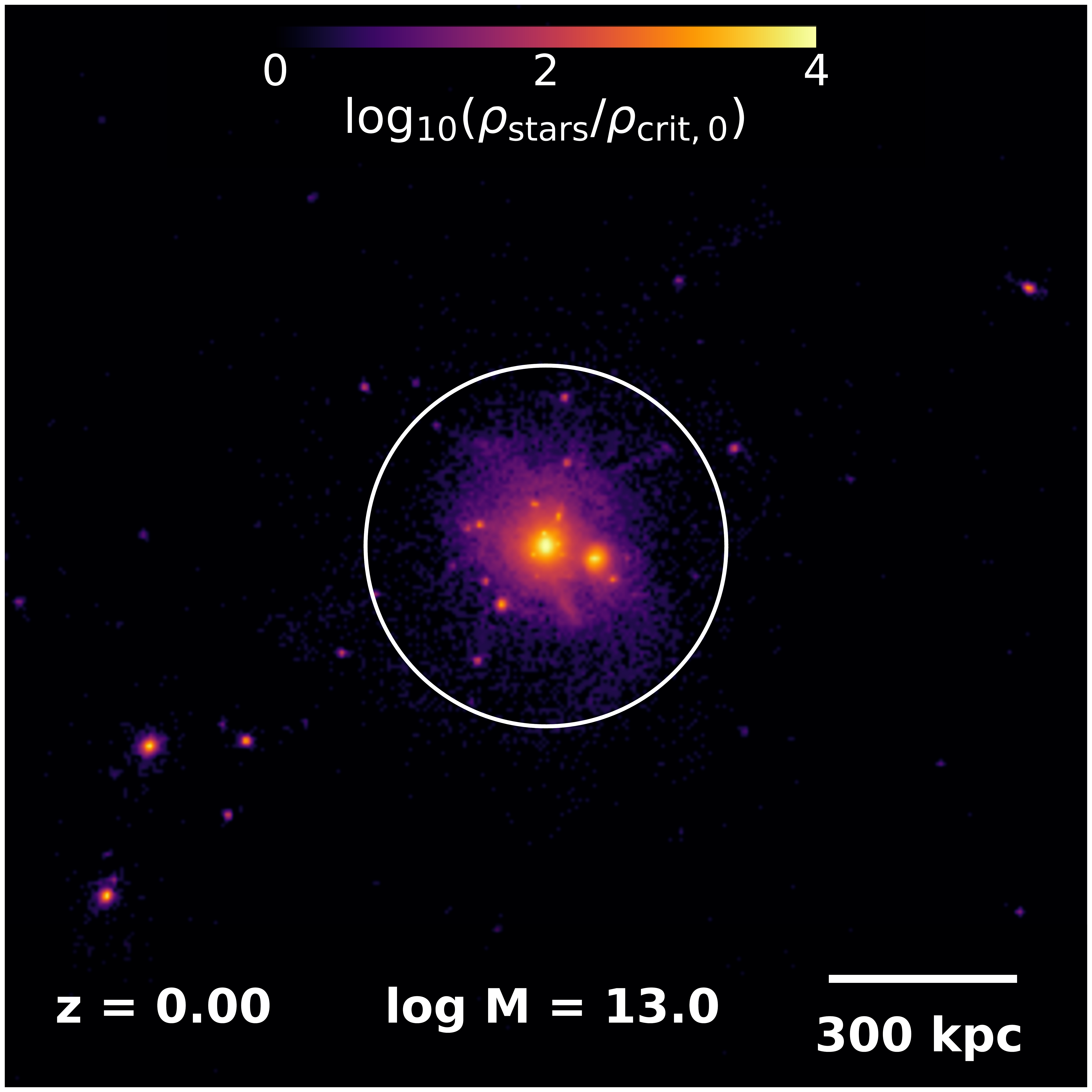}
    \includegraphics[width=0.33\textwidth]{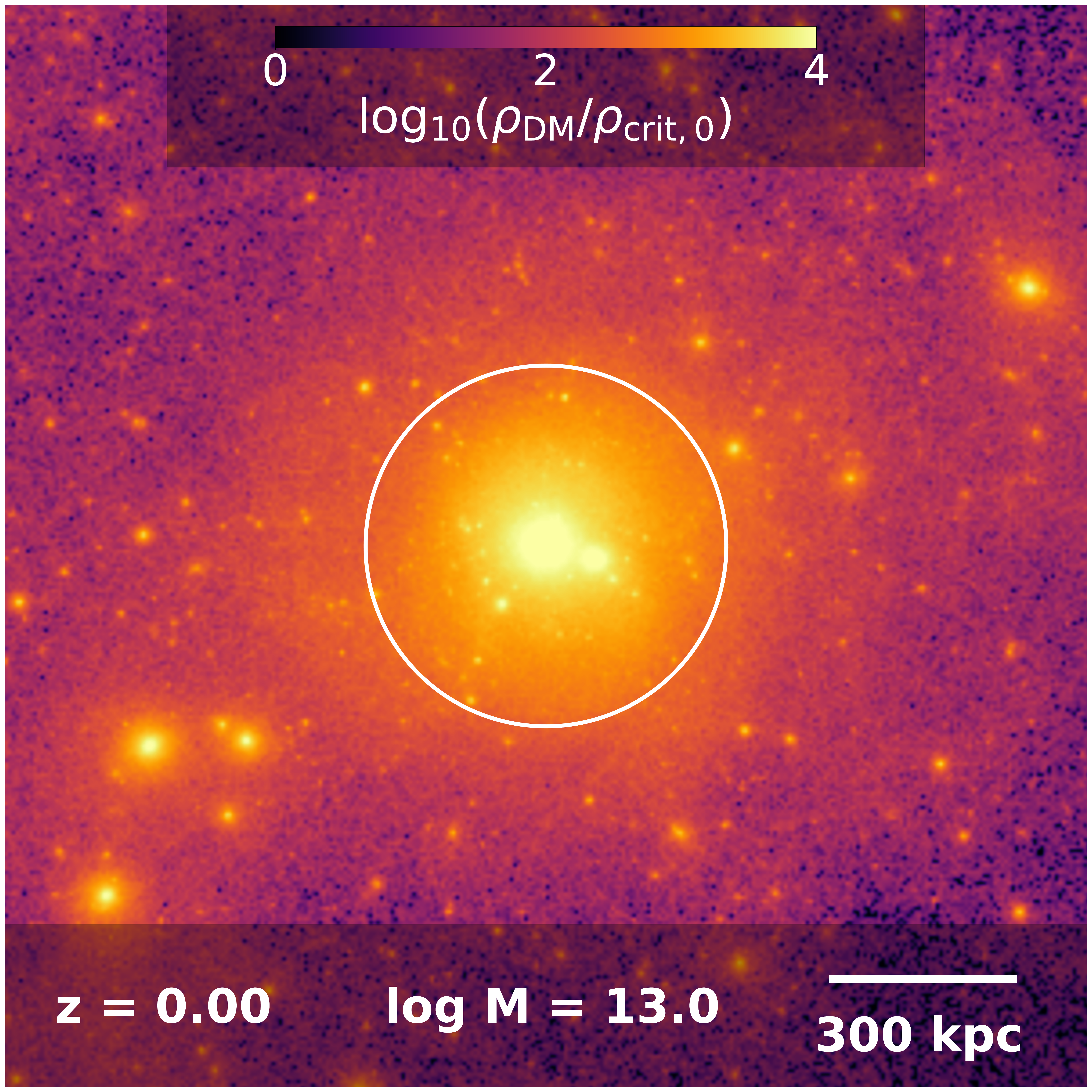}
    \caption{The distribution of baryons and dark matter in and around a halo with $M\sim 10^{13} {\rm M_{\odot}}$ at $z=0$ in the TNG100 simulation. The top panel shows the baryon fraction, normalised to the cosmic baryon fraction. Blue and red colours correspond to regions with baryon fractions lower or higher, respectively, than the cosmic value. The three smaller panels in the bottom row show the gas (left), stellar (middle), and dark matter (right) densities for the same halo, where all densities are normalised to the critical density of the Universe. In all panels, the projection depth is $2\,R_{\rm 200c}$. The baryon fraction is below the cosmic value (blue areas) within the halo boundary ($R_{\rm 200c}$, black circle) of this halo, as well as other haloes in the vicinity. In contrast, the baryon fraction is above the cosmic value (red areas) in the outskirts of the halo and in the intergalactic medium.}
\label{Fig: f_b_box}
\end{figure*}

\section{Capturing baryons in and around haloes}
\label{sec: R_c}

We begin with a visualisation of the large-scale distribution of dark matter, gas, and stars. Fig. \ref{Fig: f_b_box} illustrates the distribution of matter in and around a halo with $M\sim 10^{13} \rm M_{\odot}$ from TNG100. The large top panel shows the baryon fraction, i.e. $(M_{\rm gas} + M_{*} + M_{\rm BH})/M_{\rm tot}$, which is normalised to the cosmic baryon fraction. As a result, regions with a baryon fraction lower than the cosmic value are shown in blue, whereas regions with a baryon fraction above the cosmic value are red. The white regions contain baryon fraction equal to the cosmic value. The bottom panels depict the density of the gas (left), stars (middle), and dark matter (right), all normalised to the critical density of the Universe.

There are several noteworthy features visible. First, regions within the halo boundary ($R_{\rm 200c}$, black/white circles) are dark matter dominated (as indicated by the blue colour), while the outskirts of haloes as well as the IGM are baryon dominated (shown in red). The very centre of the large halo is red, implying that the galaxy in the innermost central region is baryon dominated. Other localised red regions with $f_{\rm b}>f_{\rm b,cosmic}$ are visible within the halo $R_{\rm 200c}$, which correspond to satellite galaxies orbiting the halo centre. These same galaxies can also be identified in the density of the stars. The other blue regions in the top panel are mostly other nearby haloes. As we will show later in this Section, almost all haloes in the TNG simulation have a baryon fraction lower than the cosmic value within their boundary. Therefore, in the framework of a $\rm \Lambda CDM$ universe, the rest of the baryons, i.e. those which are ‘missing’, must be found outside the halo, as we discuss below.

\subsection{Cumulative baryon fraction: large-scale redistribution}
\label{subsec: cumulative baryon fraction}

\begin{figure*}
    \centering
    \includegraphics[width=0.72\textwidth]{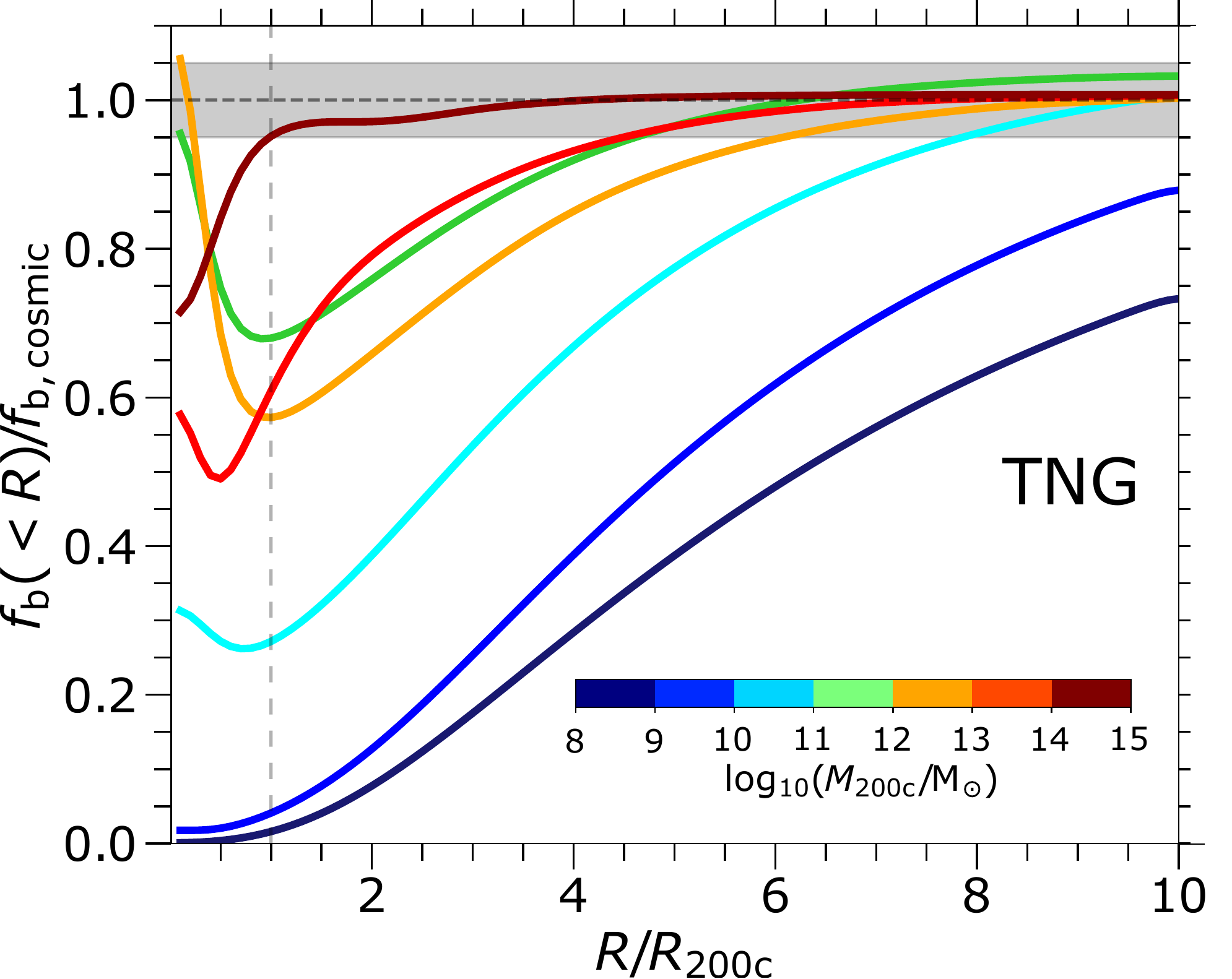}
    
    \vspace{5mm}
    
    \includegraphics[width=0.44\textwidth]{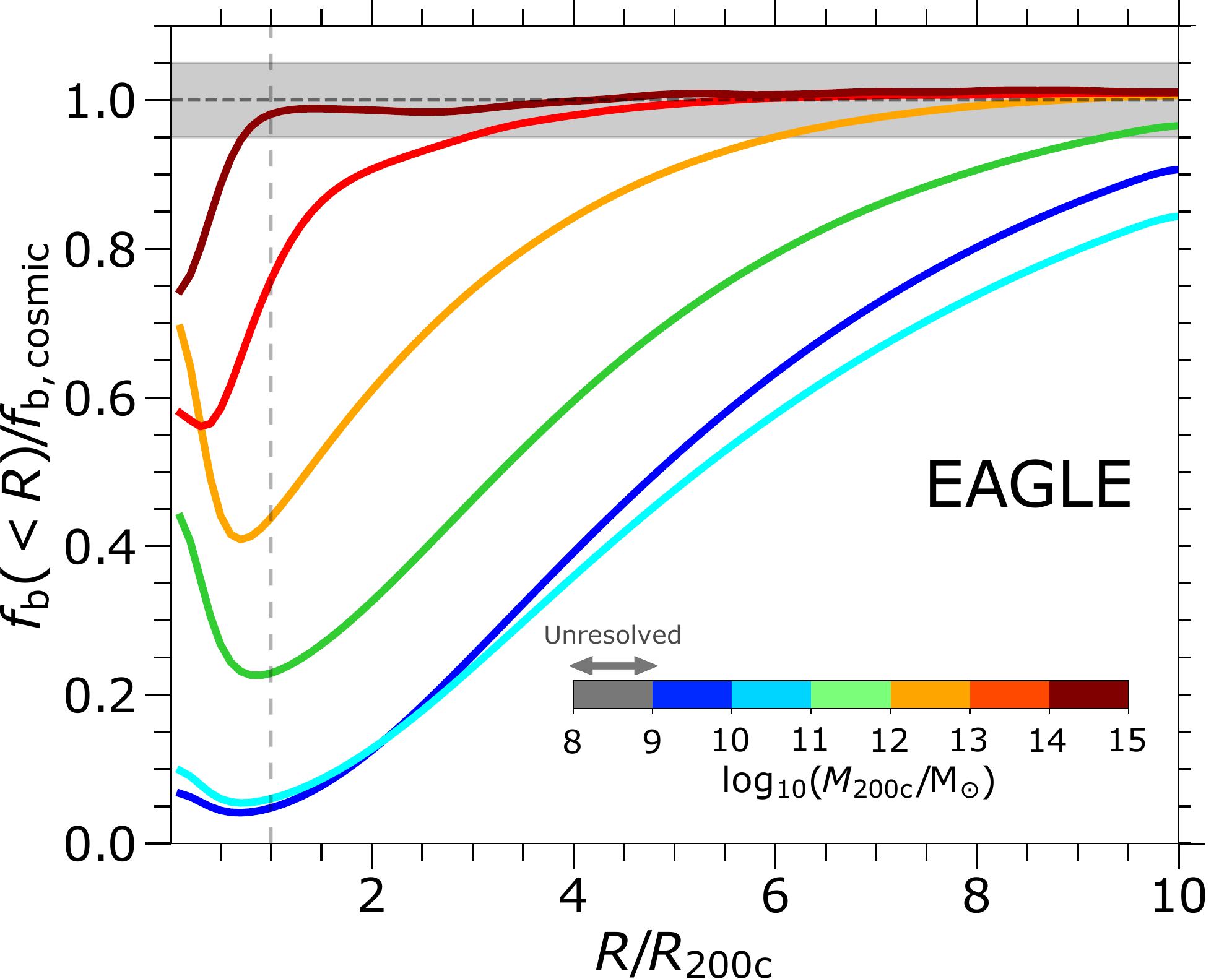}
    \includegraphics[width=0.44\textwidth]{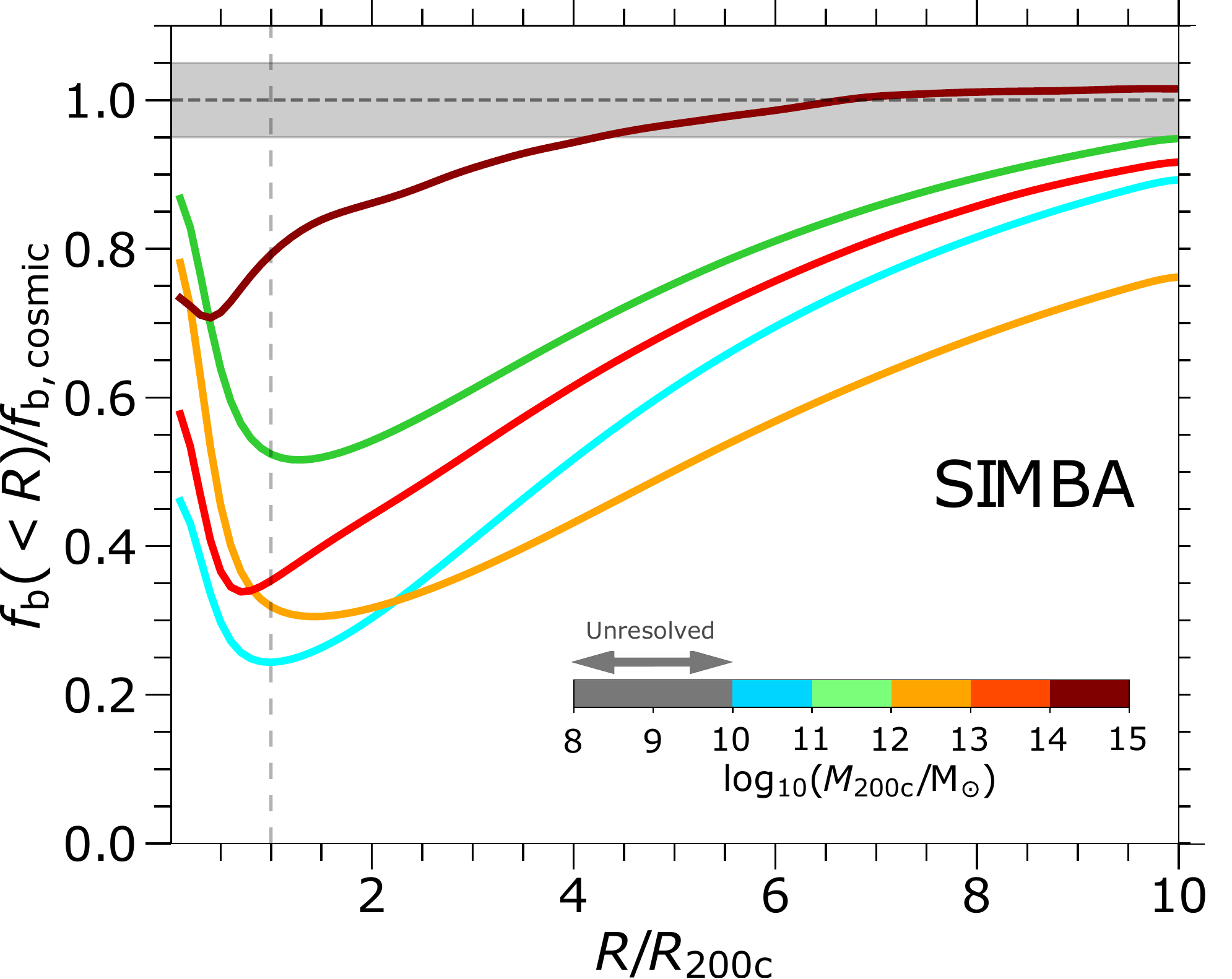}
    \caption{The profiles of mean ‘cumulative’ baryon fraction as a function of halocentric distance in TNG (top, combining TNG50, TNG100 and TNG300), EAGLE (lower left), and SIMBA (lower right) at $z=0$. The colours correspond to to discrete 1.0 dex bins of halo mass, $M_{\rm 200c}$. The dashed vertical lines show the halo boundary, $R_{\rm 200c}$. Moreover, the dashed horizontal line corresponds to the cosmic baryon fraction, and the shaded region around it depicts the observational uncertainty in the measured cosmic baryon fraction (see \citealt{planck2015_xiii}). The cumulative baryon fraction, $f_{\rm b}(<R)$, does not reach the cosmic value until far beyond the virial radius, and this trend has a complex mass dependence.
    }
\label{Fig: Profile_f_b_c}
\end{figure*} 

Fig. \ref{Fig: Profile_f_b_c} quantifies the cumulative baryon fraction, $f_{\rm b}(<R)$, normalised to the cosmic baryon fraction $f_{\rm b,cosmic}\sim 0.157$ as a function of halocentric distance. The cumulative baryon fraction is defined as the ratio between baryonic matter (gas+stars+black holes) and total matter (baryons+dark matter) within a sphere of radius $R$, centred at the halo centre. We show the results over a mass range of seven orders of magnitude, from $10^{8} \rm M_{\odot}$ to $10^{15} \rm M_{\odot}$ in the TNG (top), EAGLE (bottom left), and SIMBA (bottom right) simulations at $z=0$.\footnote{We exclude the lower mass, unresolved haloes in EAGLE and SIMBA due to a low number of particles (see Section \ref{subsubsec: FOF_SUBFIND}).} For each halo mass range, we calculate $f_{\rm b}(<R),$ by averaging over the cumulative baryon fraction of the equally weighted haloes within that particular halo mass bin.

Across the majority of the halo mass range, the average cumulative baryon fraction within $R_{\rm 200c}$ is lower than the cosmic value. The only exception is the galaxy clusters ($\log_{10}(M_{\rm 200c}/{\rm M_{\odot}})>14$), whose baryon fraction within $R_{\rm 200c}$ reaches the cosmic value in TNG and EAGLE, although not in SIMBA. This overall picture is qualitatively consistent with currently available X-ray and SZ observations (see Section \ref{subsec: obs}). The quick rise within haloes at $R/R_{\rm 200c}<0.5$, which is visible in various mass ranges and across simulations, indicates the presence of the baryon-dominated galaxy itself.

Beyond the halo boundary ($R>R_{\rm 200c}$), the cumulative baryon fraction increases with distance for all halo mass ranges until it reaches a constant value, the cosmic baryon fraction. This indicates an excess of baryons on the outskirts of haloes, compensating for the shortage of baryons within the halo boundary. Comparing different halo mass ranges (different colours) in Fig. \ref{Fig: Profile_f_b_c}, the cumulative baryon fraction is a rather complex function of the halo $M_{\rm 200c}$. For example, at $R_{\rm 200c}$ (marked by the dashed vertical line) it reaches 10-20\% of the cosmic value in dwarfs (cyan lines, $10 \lesssim \log_{10}M_{\rm 200c}/{\rm M_{\odot}} \lesssim 11$), 30-60\% in Milky Way-like haloes (orange lines, $12 \lesssim \log_{10}M_{\rm 200c}/{\rm M_{\odot}} \lesssim 13$), and 80-100\% in galaxy clusters (dark red lines, $\log_{10}M_{\rm 200c}/{\rm M_{\odot}} \gtrsim 14$), depending on the simulation. This shows that haloes of different masses are subject to different physical processes.

\subsection{The closure radius: a new characteristic scale}
\label{subsec: closure_radius}

By construction, the baryon fraction within every $\rm \Lambda CDM$ cosmological simulation box equals the cosmic value in the initial conditions.
Consequently, for every halo in a $\rm \Lambda CDM$ universe, there \textit{must} be a characteristic scale, the closure radius $R_{\rm c}$, within which the cumulative baryon fraction equals the cosmic baryon fraction. In other words, $R_{\rm c}$ is the halocentric distance within which we can find all missing baryons,
\begin{equation}
    f_{\rm b}(<R_{\rm c}) = f_{\rm b,cosmic} \pm \Delta f_{\rm b,cosmic}.
\end{equation}
We adopt $\Delta f_{\rm b,cosmic} \sim 0.05 f_{\rm b,cosmic}$ from the observational uncertainty on the cosmic baryon fraction \citep{planck2015_xiii}.

This characteristic radius can already be inferred from Fig. \ref{Fig: Profile_f_b_c} for halo masses where $f_{\rm b}(<R)$ reaches the cosmic value within the dynamical range of the x-axis of the plot. For reference, in clusters with $\log_{10}(M_{\rm 200c}/{\rm M_{\odot}})\gtrsim14$ (dark red lines), $R_{\rm c}$ is around $R_{\rm 200c}$ in TNG, $0.5 R_{\rm 200c}$ in EAGLE, and $4 R_{\rm 200c}$ in SIMBA.

We measure $R_{c}$ for all the haloes in TNG, EAGLE, and SIMBA, and show the result in the top panel of Fig. \ref{Fig: R_c}, which plots the median of the normalised closure radius $R_{c}/R_{\rm 200c}$ (solid lines) as a function of the halo mass.\footnote{We measure and discuss the closure radius for the original Illustris simulation in Appendix \ref{app: Illustris}.} Moreover, the bottom panel of Fig. \ref{Fig: R_c} shows 2D histograms of the cumulative baryon fraction as a function of $M_{\rm 200c}$ (y-axis) and distance (x-axis).
The arrows show the mass ranges where a given physical process is the most dominant factor in reshaping the distribution of baryons and determining the amplitude of $R_{\rm c}$, as discussed below.

The amplitude of the closure radius changes with the halo mass. As it can be seen in the top panel of Fig. \ref{Fig: R_c}, $R_{\rm c}/R_{\rm 200c}$ is large at low halo masses and decreases with halo mass. In TNG and SIMBA, a rise occurs at $12\lesssim \log_{10}(M_{\rm 200c}/{\rm M_{\odot}})\lesssim 13$ due to their particular models for AGN feedback (see discussion below). On the other hand, $R_{\rm c}/R_{\rm 200c}$ decreases monotonically with the halo mass in EAGLE, which has a different, and non-kinetic, model for AGN feedback. In all three simulations, the closure radius becomes equal to or smaller than the halo $R_{\rm 200c}$ in galaxy clusters. We emphasise that although haloes reach the cosmic value at the closure radius, the region within the closure radius should not be considered a closed box \citep[see][]{Mitchell2022Baryonic}. Indeed, inflows and outflows may still alter the properties of haloes and influence their evolution. In future work, we study the kinematics of gas flows in and around haloes.

\begin{figure*}
    \centering
    \includegraphics[width=0.78\textwidth]{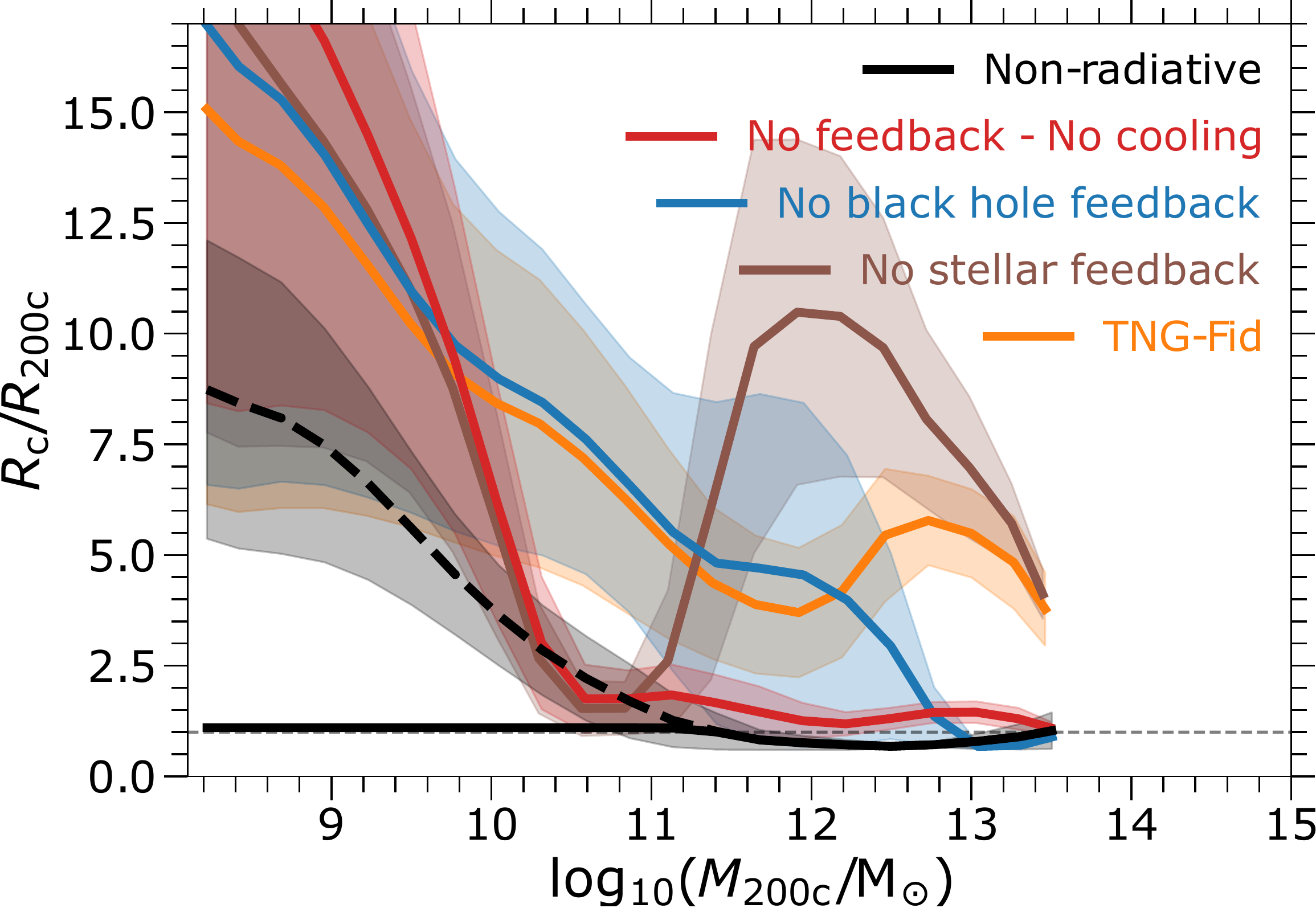}
    
    \vspace{5mm}
    
    \includegraphics[width=0.85\columnwidth]{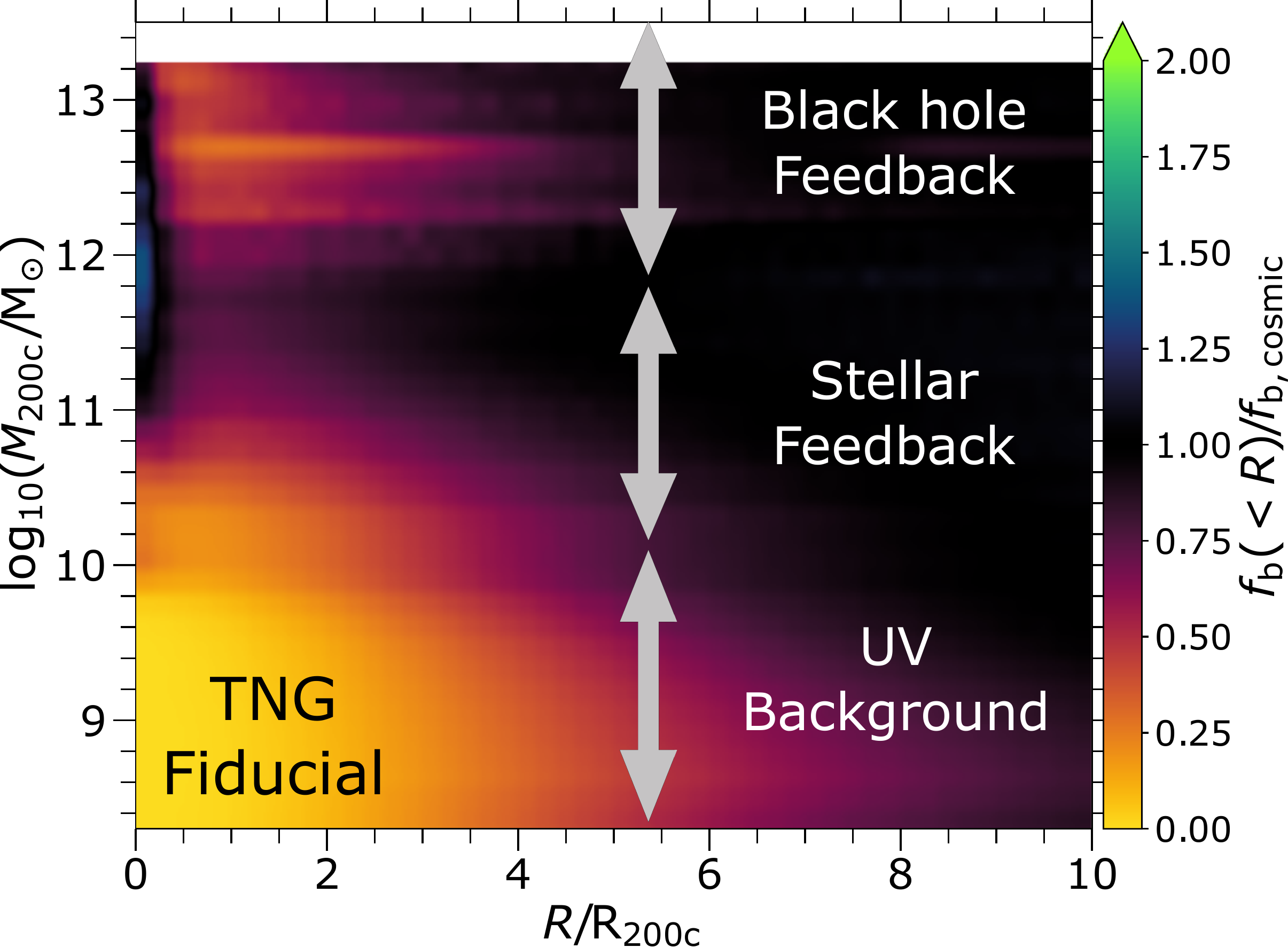}
    \hspace{3mm}
    \includegraphics[width=0.85\columnwidth]{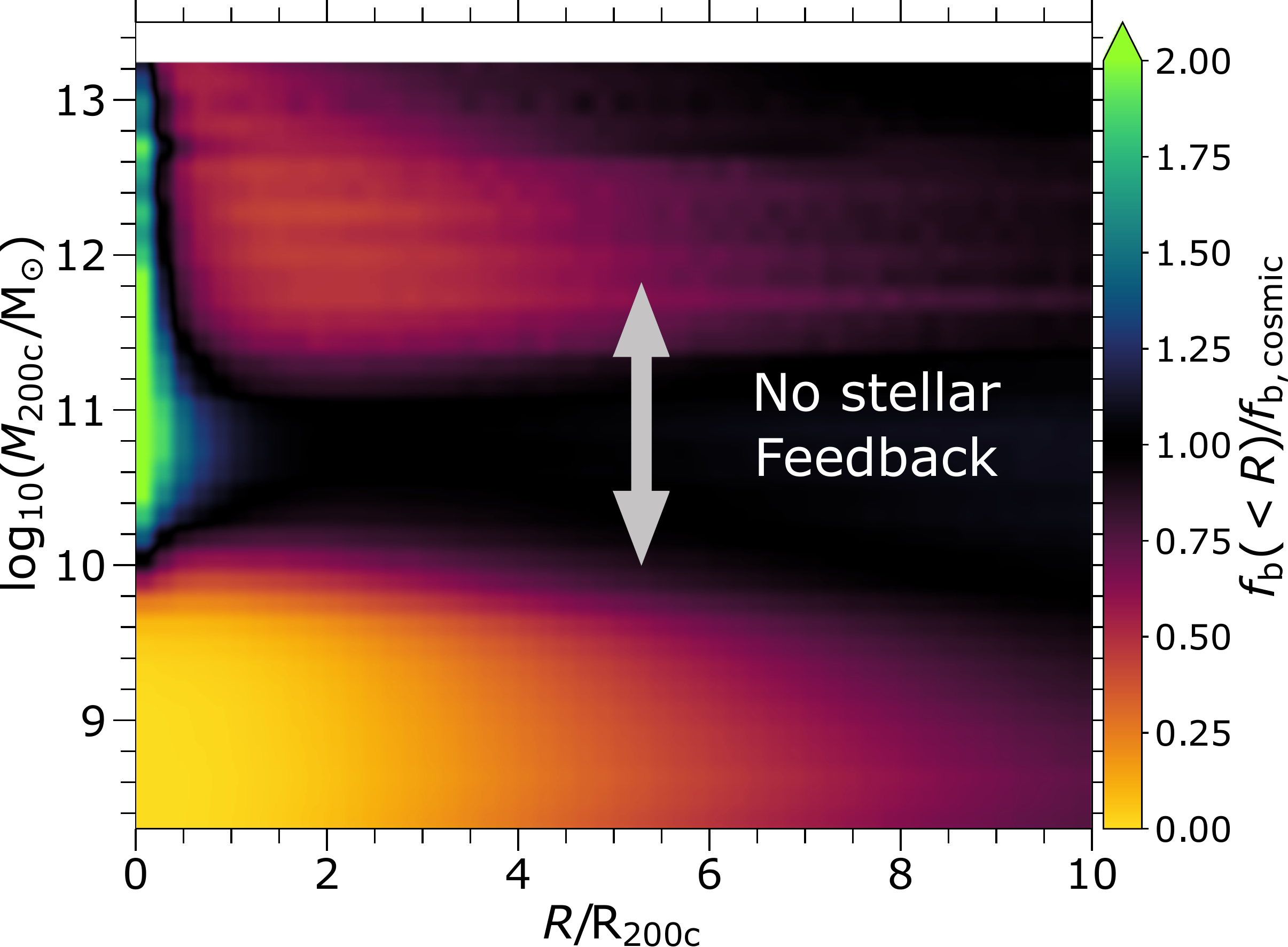}
    \includegraphics[width=0.85\columnwidth]{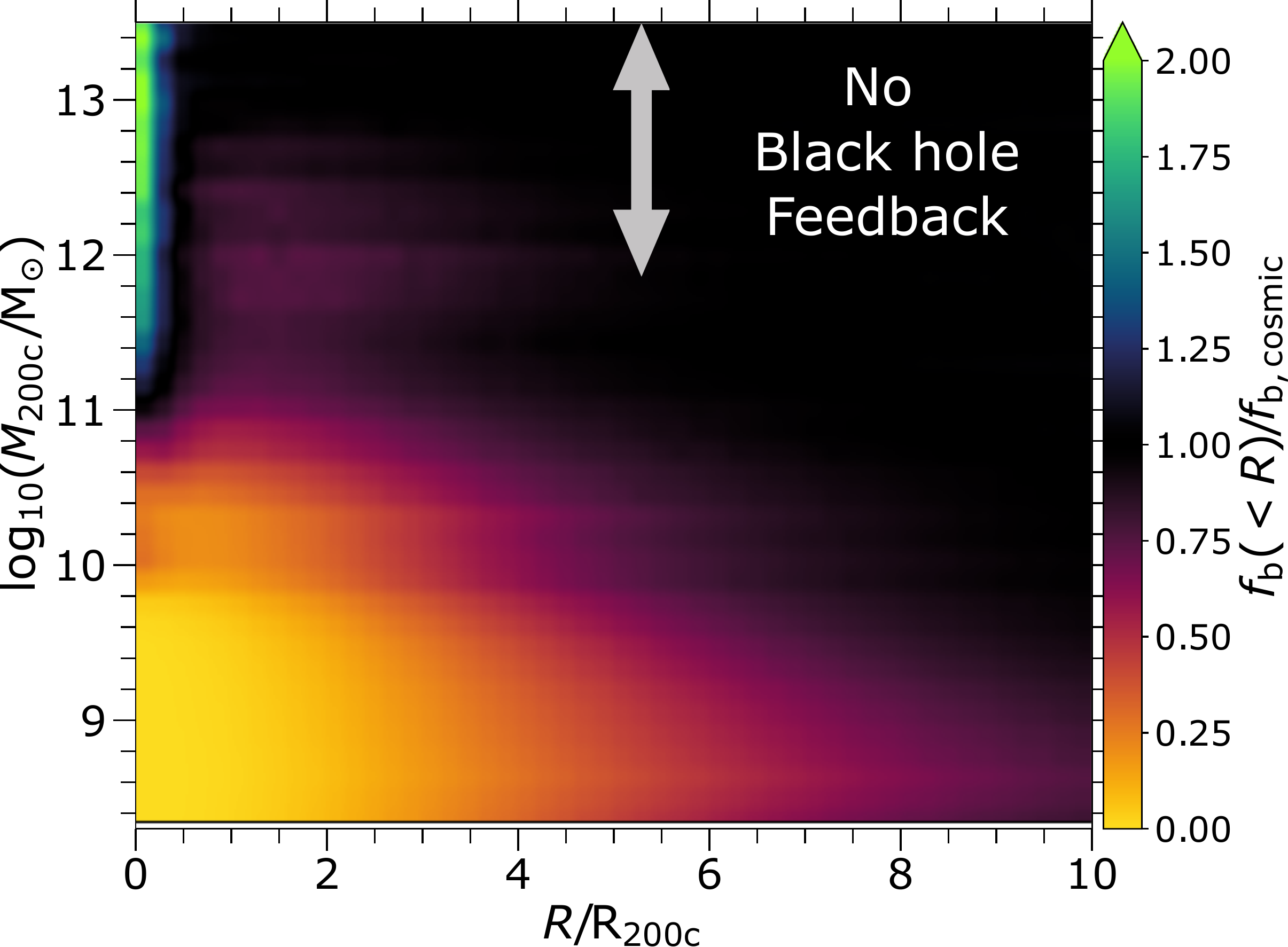}
    \hspace{3mm}
    \includegraphics[width=0.85\columnwidth]{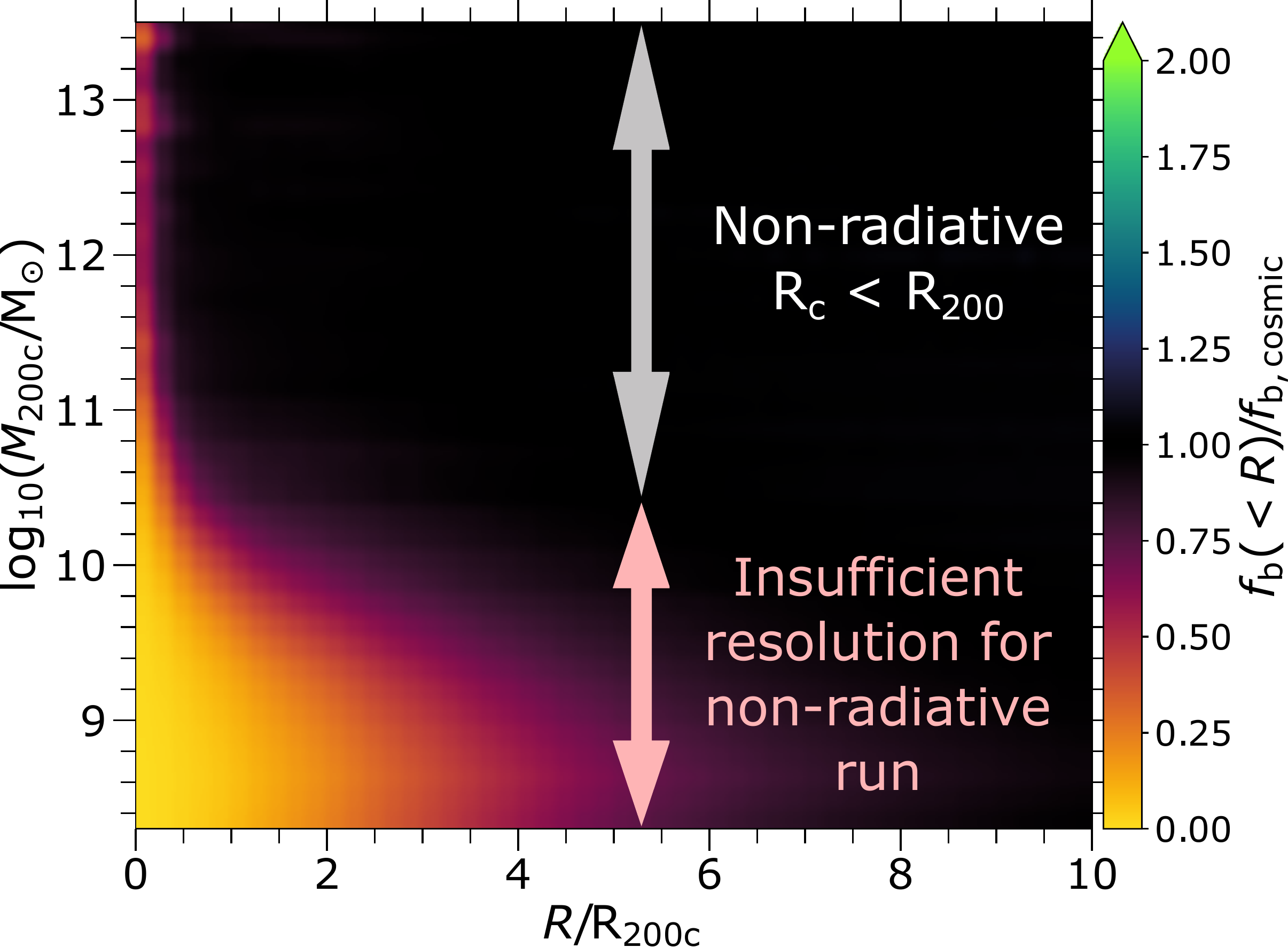}
    \caption{Top row: The characteristic scale i.e. ``closure radius'' $R_{\rm c}$ within which all baryons associated with dark matter are found, for several variants of the TNG model performed in $\sim \rm 37 Mpc$ boxes (see text). We show the median value as a function of halo $M_{\rm 200c}$ over all haloes at each given $M_{\rm 200c}$ bin. The shaded regions correspond to 16\% and 84\% percentiles. The solid black line corresponds to our expectation of the non-radiative run at a sufficiently high-resolution, while the dashed line is the actual resolution-impacted result at the resolution of the other depicted runs (see Appendix \ref{app: TNG_convergence}). Middle and bottom rows: 2D histograms of the average ‘cumulative’ baryon fraction (colours) as a function of the halo mass (y-axis) and halocentric distance (x-axis) in four TNG model variants. The orange and green colours correspond to the cumulative baryon fraction less and greater than the cosmic value, respectively. We can see that $R_{\rm c}>R_{\rm 200c}$, i.e. the excess of baryons beyond $R_{\rm 200c}$, is explained by the influence of astrophysical feedback processes, each of which impacts $R_{\rm c}$ at different halo mass scales.}
\label{Fig: TNG_VAR_R_c}
\end{figure*}

To understand the complex behaviour of $R_{c}$, and to investigate the impact of different physical processes on the distribution of baryons and the amplitude of $R_{\rm c}$, we turn to an analysis of five specific variation simulations of the TNG model, out of the $\sim 100$ available \citep[see][]{nelson2018ovi,pillepich2018First}. These runs are: with the fiducial TNG model, with no stellar feedback, with no black holes, with no feedback and no cooling, and non-radiative (see Section \ref{subsubsec: TNG}). In the top row of Fig. \ref{Fig: TNG_VAR_R_c}, we show $R_{\rm c}$ as a function of halo mass in these variants. We note that the trends are available only until $\log_{10}(M_{\rm 200c}/{\rm M_{\odot}}) \sim 13$ because these smaller volumes do not sample the high-mass end of the halo mass function. In the middle and bottom rows, we also include 2D histograms of baryon fraction, as in Fig. \ref{Fig: R_c}. Comparing our findings in different simulations, shown in Figs. \ref{Fig: R_c} and \ref{Fig: TNG_VAR_R_c}, we explain the behaviour of $R_{\rm c}$ as a function of the halo mass by dividing $M_{\rm 200c}$ into four regions:

\begin{figure*}
    \centering
    \includegraphics[width=0.8\textwidth]{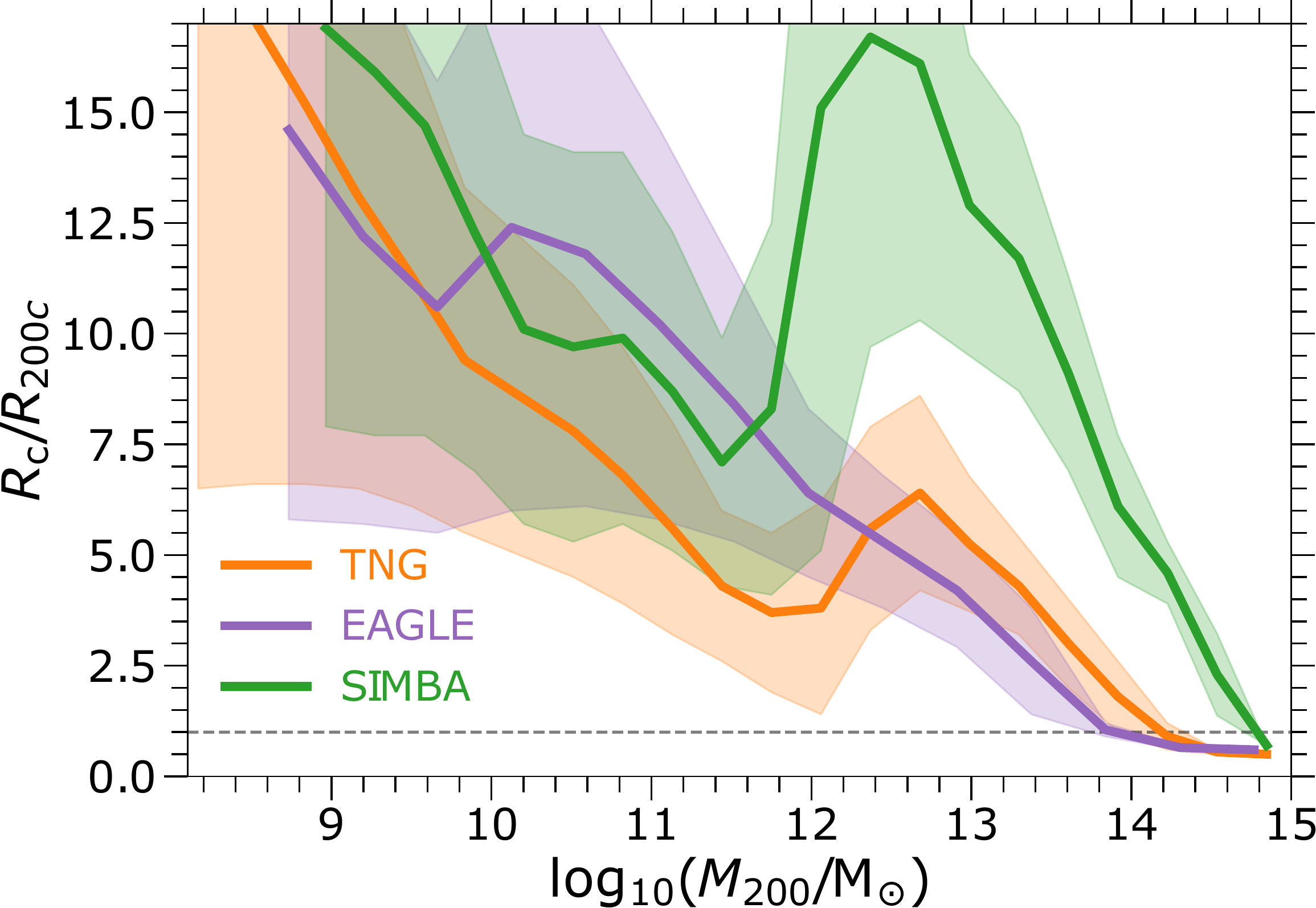}
    
    \vspace{6mm}
    
    \includegraphics[width=1\textwidth]{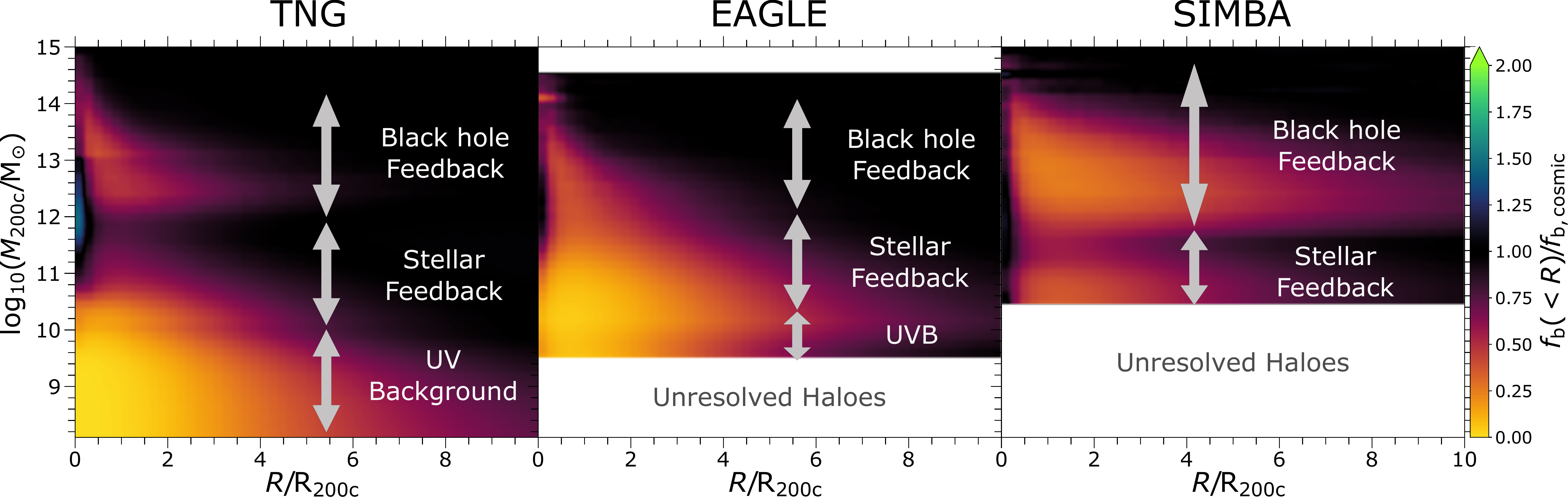}
    \caption{Top panel: The characteristic scale i.e. ``closure radius'' $R_{\rm c}$ within which all baryons associated with dark matter are found in TNG, EAGLE, and SIMBA at $z=0$. We show the median value as a function of halo $M_{\rm 200c}$ over all haloes at each given $M_{\rm 200c}$ bin. The shaded regions correspond to 16\% and 84\% percentile halo-to-halo variation. The closure radius is a strong function of the halo mass with different amplitude for each simulation. Bottom row: 2D histograms of the average ‘cumulative’ baryon fraction (colours) as a function of the halo mass (y-axis) and halocentric distance (x-axis) in TNG, EAGLE, and SIMBA. The orange bright regions show regimes where baryons are missing, whereas the dark regions show regions at the cosmic baryon fraction.
    }
\label{Fig: R_c}
\end{figure*}

\begin{figure*}
    \centering
    \includegraphics[width=0.7\textwidth]{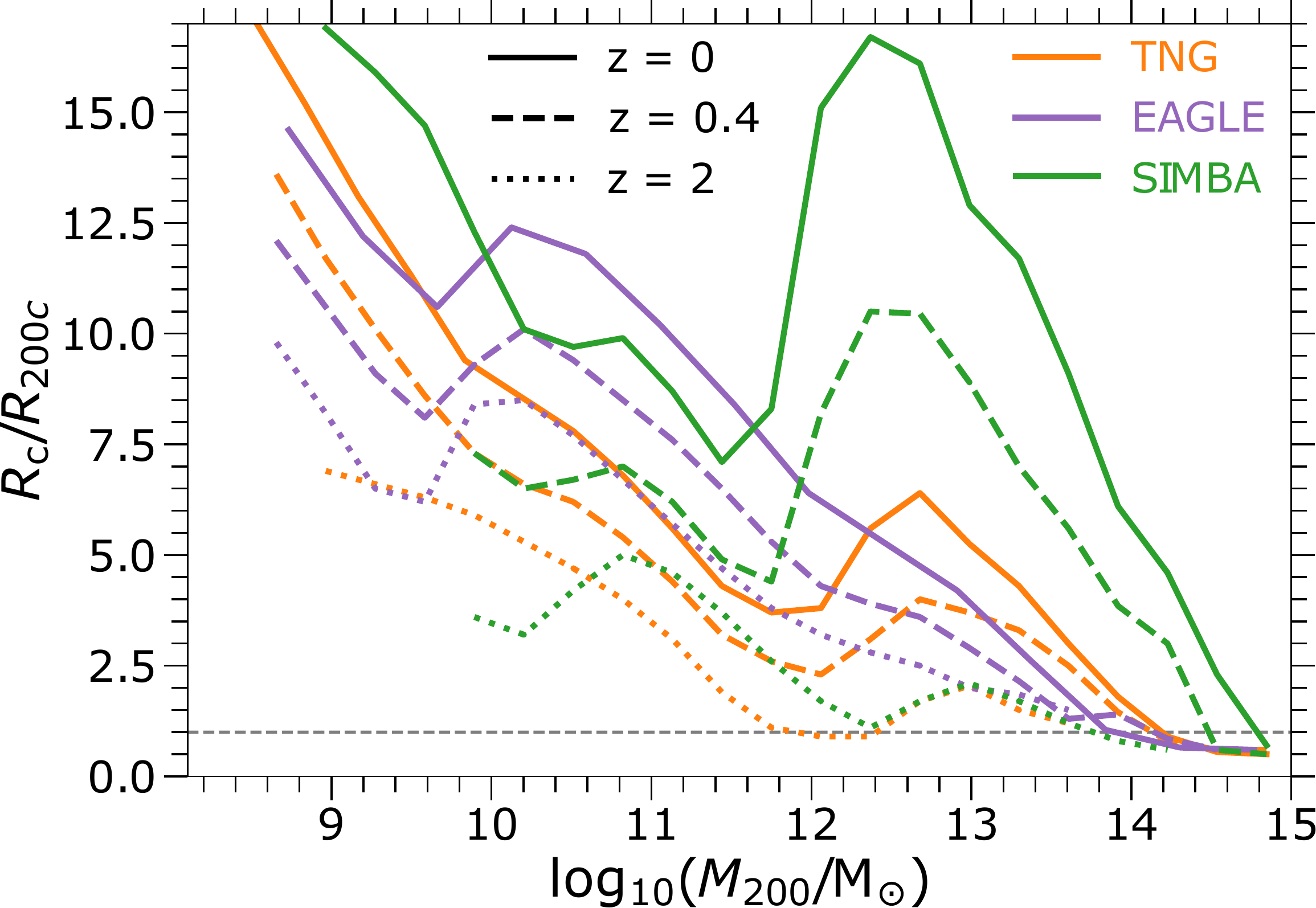}
    \caption{The closure radius as a function of $M_{200}$ in TNG, EAGLE, and SIMBA at different cosmic epochs. Different linestyles show different redshifts, namely $z=0$ (solid), an intermediate look-back time at $z=0.4$ (dashed), and the early Universe at $z=2$ (dotted). The qualitative behavior of closure radius with mass is in place even at $z=2$. However, at fixed halo mass, the value of $R_{\rm c}$ always decreases with redshift in all three models. At the population level, baryon ejection to large-scales is a gradual process.}
\label{Fig: R_c_z}
\end{figure*}

\begin{itemize}
    \item Dark haloes ($8 \lesssim \log_{10}M_{\rm 200c}/{\rm M_{\odot}}\lesssim 10$): For ultra low-mass dark matter haloes, UV background radiation is the major physical process preventing infall of the gas into haloes (Fig. \ref{Fig: R_c}). These haloes do not contain any luminous galaxies. Their potential wells are shallow so they are unable to accrete and keep gas from the intergalactic medium. Consequently, $R_{\rm c}/R_{\rm 200c}$ in these haloes is relatively large. The non-radiative run of TNG (Fig. \ref{Fig: TNG_VAR_R_c}) shows that the amplitude of $R_{\rm c}$ in this case is smaller than the fiducial model, and than any other variant. It would reach $R_{\rm 200c}$ in the limit of very high resolution (black solid line; see Appendix \ref{app: TNG_convergence}).
    \item Haloes hosting low-mass galaxies ($10 \lesssim \log_{10}M_{\rm 200c}/{\rm M_{\odot}}\lesssim 12$): Comparing $R_{\rm c}$ in TNG fiducial versus the simulation without stellar feedback (top panel of Fig. \ref{Fig: TNG_VAR_R_c}), we find that stellar feedback is the dominant physical process influencing $R_{\rm c}$ and the distribution of baryons in this halo mass range. This is also true for EAGLE and SIMBA (Fig. \ref{Fig: R_c}), although the quantitative mass range where stellar feedback dominates differs by $\pm \rm 0.5 \,dex $. The characteristic $R_{\rm c}$ decreases with the halo mass until it reaches $R_{\rm c} \sim R_{\rm 200c}$ at $\log_{10}(M_{\rm 200c}/{\rm M_{\odot}}) \sim 11$ in the TNG run with no stellar feedback, as opposed to $R_{\rm c} \sim 5 R_{\rm 200c}$ in the fiducial TNG model.
    
    An important caveat exists for the no stellar feedback simulation. Due to runaway cooling in low-mass haloes, these galaxies undergo rapid stellar mass growth, as well as supermassive black hole growth, such that the inner part of the haloes becomes baryon-dominated (Fig. \ref{Fig: TNG_VAR_R_c}, middle right panel). More importantly, strong, and ejective, AGN feedback sets in at $\log_{10}(M_{\rm 200c}/{\rm M_{\odot}}) \gtrsim 11$ in contrast to the fiducial run where this happens at $\log_{10}(M_{\rm 200c}/{\rm M_{\odot}}) \gtrsim 12$. This leads to an unphysically stronger AGN feedback, more ejected gas, and, therefore, larger $R_{\rm c}$ in this mass range. In addition, comparing the run with no black hole feedback to the fiducial model, we do not find any significant impact from thermal black hole feedback\footnote{The thermal black hole feedback is, in general, the active mode for $z=0$ TNG haloes with $\log_{10}(M_{\rm 200c}/{\rm M_{\odot}}) \lesssim 12$ (see Section \ref{subsubsec: AGN_feedback})} on $R_{\rm c}$ or on the cumulative baryon fraction.
    
    \item Milky Way to group-mass haloes ($12 \lesssim \log_{10}M_{\rm 200c}/{\rm M_{\odot}}\lesssim 14$): Supermassive black hole feedback is the dominant physical process reshaping the baryonic distribution and influencing $R_{\rm c}$ in this mass range. In TNG and SIMBA, a significant increase in $R_{\rm c}$ at $\log_{10}(M_{\rm 200c}/{\rm M_{\odot}})\sim 12$ is obvious (Fig. \ref{Fig: R_c}), which occurs due to AGN feedback. This is clear in the comparison with a TNG variant with no black holes. In this simulation, $R_{\rm c}$ does not increase at $\log_{10}(M_{\rm 200c}/{\rm M_{\odot}}) \sim 12$, but continues monotonically decreasing until it reaches $R_{\rm 200c}$ at $\log_{10}(M_{\rm 200c}/{\rm M_{\odot}})\sim 13$ (Fig. \ref{Fig: TNG_VAR_R_c}, top panel). This shows that the kinetic mode of AGN feedback ejects gas beyond the halo boundary, impacting baryons out to relatively large scales. This kinetic feedback is significantly stronger in SIMBA, such that $R_{\rm c}$ is substantially larger than in TNG for this halo mass range. On the other hand, $R_{\rm c}$ decreases monotonically with the halo mass in EAGLE, which does not implement a switch between more than one distinct AGN feedback mode. For reference, at $\log_{10}(M_{\rm 200c}/{\rm M_{\odot}}) \sim 12.5$, the closure radius is $16\, R_{\rm 200c}$ in SIMBA, $6\, R_{\rm 200c}$ in TNG, $5\, R_{\rm 200c}$ in EAGLE, and $2\rm R_{\rm 200c}$ in TNG with no black holes.
    
    \item Galaxy clusters ($\log_{10}M_{\rm 200c}/{\rm M_{\odot}}\gtrsim 14$): Neither stellar feedback nor supermassive black hole feedback substantially influences $R_{\rm c}$ or the distribution of baryons in these massive haloes in TNG and EAGLE. Their deep gravitational potential wells and large sizes do not allow efficient gas ejection beyond the halo boundary. Therefore, $R_{\rm c}\lesssim R_{\rm 200c}$ in these two simulations. On the other hand, the strong black hole feedback in SIMBA still ejects gas beyond $R_{\rm 200c}$ of clusters, and $R_{\rm c} \sim R_{\rm 200c}$ only for the most massive clusters with $\log_{10}(M_{\rm 200c}/{\rm M_{\odot}}) \sim 15$.
\end{itemize}

\subsection{The closure radius at different cosmic epochs}
\label{subsec: R_c_time_ev}

The closure radius is a time-dependent quantity. In Fig. \ref{Fig: R_c_z} we show the normalised closure radius, $R_{\rm c}/R_{\rm 200c}$ as a function of $M_{\rm 200c}$ at three different redshifts (different linestyles) in the TNG, EAGLE, and SIMBA simulations. This does not represent the time evolution of the closure radius as individual haloes evolve, but rather the average closure radius of haloes of a given mass at different cosmic epochs.

At fixed halo mass, the amplitude of $R_{\rm c}/R_{\rm 200c}$ always decreases with redshift in all three simulations. We confirm this statement based on our analysis of different simulation snapshots, three of which we show in Fig. \ref{Fig: R_c_z}. In TNG and SIMBA, AGN feedback causes a peak in the closure radius at $\log_{10}(M_{\rm 200c}/{\rm M_{\odot}}\sim 13)$ at all redshfits, although the exact halo mass of this peak slightly varies with redshift. On the other hand, $R_{\rm c}$ for EAGLE remains monotonically decreasing with the halo mass for $\log_{10}(M_{\rm 200c}/{\rm M_{\odot}}> 10)$ at all times.

Overall, simulations are in better agreement with each other at high redshift. This shows that the integrated impact of physical processes on the distribution of baryons increases with time.\footnote{We note that the trends of the closure radius with time are qualitatively similar, and non-vanishing, if we normalise $R_{\rm c}$ by $R_{\rm 200m}$, the radius within which the density is 200 times the mean density, instead of $R_{\rm 200c}$.}

\subsection{Correlations between the closure radius and galaxy properties}
\label{subsec: correlation_galprop}
\begin{figure*}
    \centering
    \includegraphics[width=1\columnwidth]{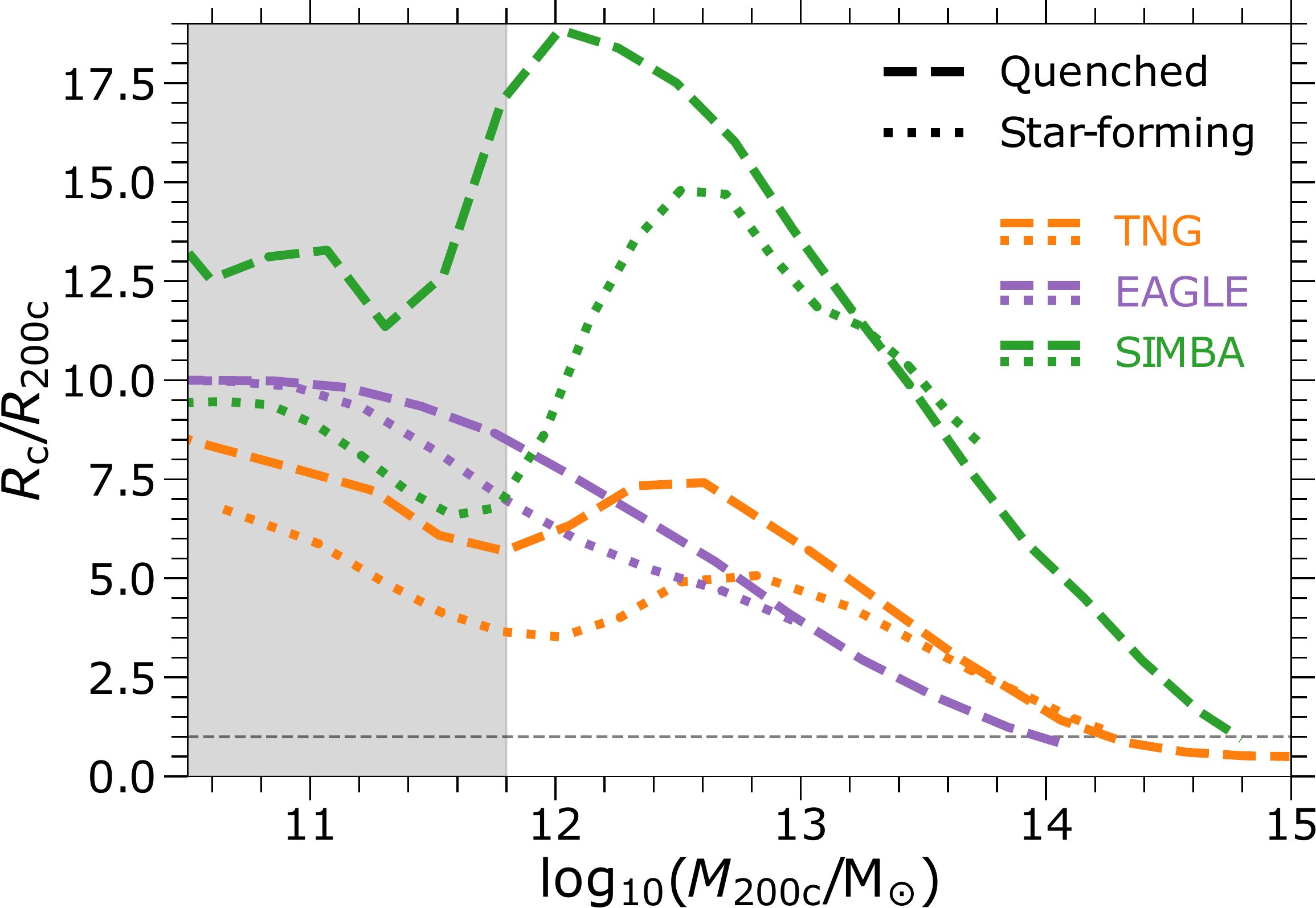}
    \includegraphics[width=1\columnwidth]{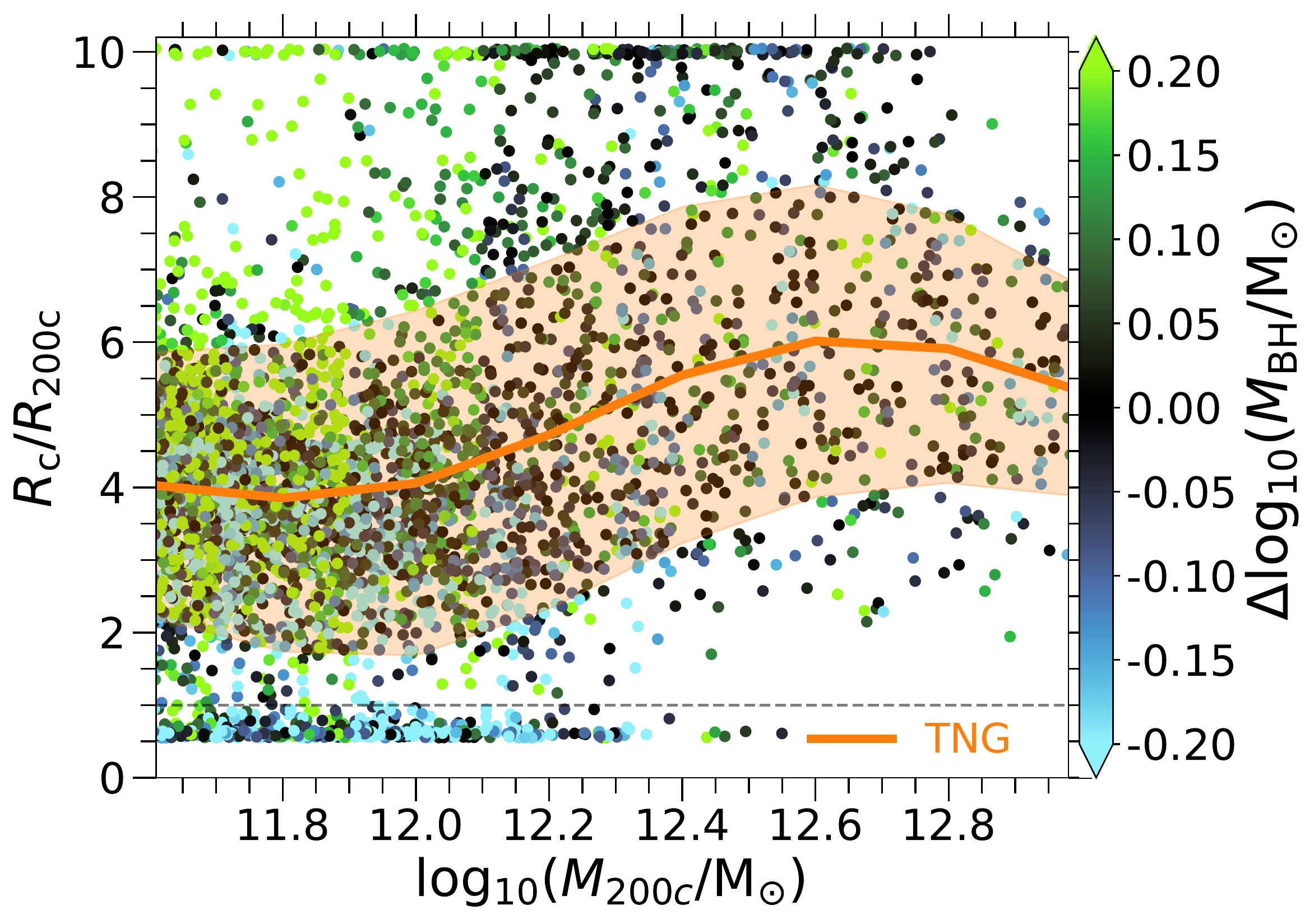}
    \caption{The correlation between the closure radius and galaxy properties. Left: The characteristic closure radius $R_{\rm c}$, splitting the galaxy population into star-forming (dotted) and quenched (dashed) sub-samples, based on a sSFR threshold of $\rm sSFR = 10^{-11} yr^{-1}$. We compare the TNG, EAGLE, and SIMBA simulations at $z=0$.
    We show the median value as a function of halo $M_{\rm 200c}$. The difference between $R_{\rm c}$ in quenched and star-forming haloes is significant in TNG and SIMBA, and much smaller in EAGLE. Right: $R_{\rm c}/R_{\rm 200c}$ (y-axis) as a function of halo mass (x-axis). Each dot corresponds to a single halo, coloured by the logarithm of the ratio between its black hole mass and the median black hole mass of all haloes in the same halo mass bin. Haloes with $M_{\rm 200c}/{\rm M_{\odot}}\lesssim 12.5$ which host undermassive black holes have significantly smaller closure radii, which reflects their AGN feedback activity.}
\label{Fig: R_c_galprop}
\end{figure*}

In Fig. \ref{Fig: R_c_galprop} we investigate the correlation between $R_{\rm c}$ and different properties of galaxies in TNG, EAGLE, and SIMBA. The left panel shows $R_{\rm c}/R_{\rm 200c}$ as a function of the halo mass for quenched (dashed lines) and star-forming (dotted lines) haloes at $z=0$. Here we consider a halo quenched if its total specific star formation rate (${\rm sSFR} = {\rm SFR}_{\rm halo}/M_{\rm \star, halo}$) is below $10^{-11} \rm yr^{-1}$. This threshold is motivated by several observational and theoretical studies \citep[e.g. see][]{Wetzel2012Galaxy,henriques17}. We also note that low-mass central galaxies ($\log_{10}M_{\rm 200c}/{\rm M_{\odot}}\lesssim 11.8$, shaded region) are typically star-forming \citep[e.g.][]{donnari19}. The quenched centrals in this mass range are predominantly backsplash or fly-by galaxies, which are currently centrals but have been satellites of other hosts at higher redshifts \citep[e.g.][]{donnari2020b}. The distribution of baryons around these galaxies would have been severely altered by environmental effects in the past \citep[][]{ayromlou2019new,Ayromlou2021Galaxy,Borrow2022There}. Therefore, we focus our analysis on haloes of $\log_{10}(M_{\rm 200c}/{\rm M_{\odot}})\gtrsim 11.8$ to avoid mixing up physical effects. However, we keep lower mass haloes in Fig. \ref{Fig: R_c_galprop} for completeness.

Overall, quenched galaxies have systematically larger $R_{\rm c}$ in all three models. The difference between $R_{\rm c}/R_{\rm 200c}$ in quenched and star-forming galaxies in TNG and SIMBA is larger than in EAGLE. Although AGN feedback is the primary mechanism for quenching galaxies in all three models, the kinetic mode of AGN feedback in TNG and SIMBA appears to have a stronger ejective character \citep[see][]{Mitchell2020Galactic}, consequently increasing $R_{\rm c}$ to a greater degree. This difference is maximal when AGN feedback is most efficient ($\log_{10}M_{\rm 200c}/{\rm M_{\odot}}\sim 12-13$), but becomes negligible for more massive haloes.

The right panel of Fig. \ref{Fig: R_c_galprop} shows $R_{\rm c}/R_{\rm 200c}$ as a function of $M_{\rm 200c}$ (x-axis) for TNG. Each circle indicates a single halo, and is coloured according to SMBH mass \textit{relative} to the median SMBH mass at that halo mass. We see that $R_{\rm c}/R_{\rm 200c}$ shows a clear trend with the black hole mass: at fixed halo mass, haloes with undermassive black holes (bluer dots) typically have lower $R_{c}/R_{\rm 200c}$ than haloes with overmassive black holes (greener dots). This is true for haloes with $11.5<\log_{10}(M_{\rm 200c}/{\rm M_{\odot}})<12.5$. The main difference between undermassive and overmassive SMBHs is that the former are still in the thermal mode, whereas the latter have switched to the kinetic mode. This supports our previous findings that AGN feedback is the main process reshaping the distribution of baryons in and around haloes at this mass scale. Above a certain halo mass $\log_{10}(M_{\rm 200c}/{\rm M_{\odot}})\gtrsim 12$, most black holes, even relatively undermassive ones, are sufficiently massive to have activated the kinetic mode of AGN feedback, thereby eliminating this tertiary trend. We also looked for a connection between $R_{\rm c}/R_{\rm 200c}$ and the galaxy stellar mass at fixed halo mass, but did not find any significant correlation.


\section{Discussion and in-depth considerations}
\label{sec: discussion}

\subsection{A universal relation between the closure radius and the halo baryon fraction}
\label{subsec: universal_relation_R_c}

\begin{figure*}
    \centering
    \includegraphics[width=0.99\columnwidth]{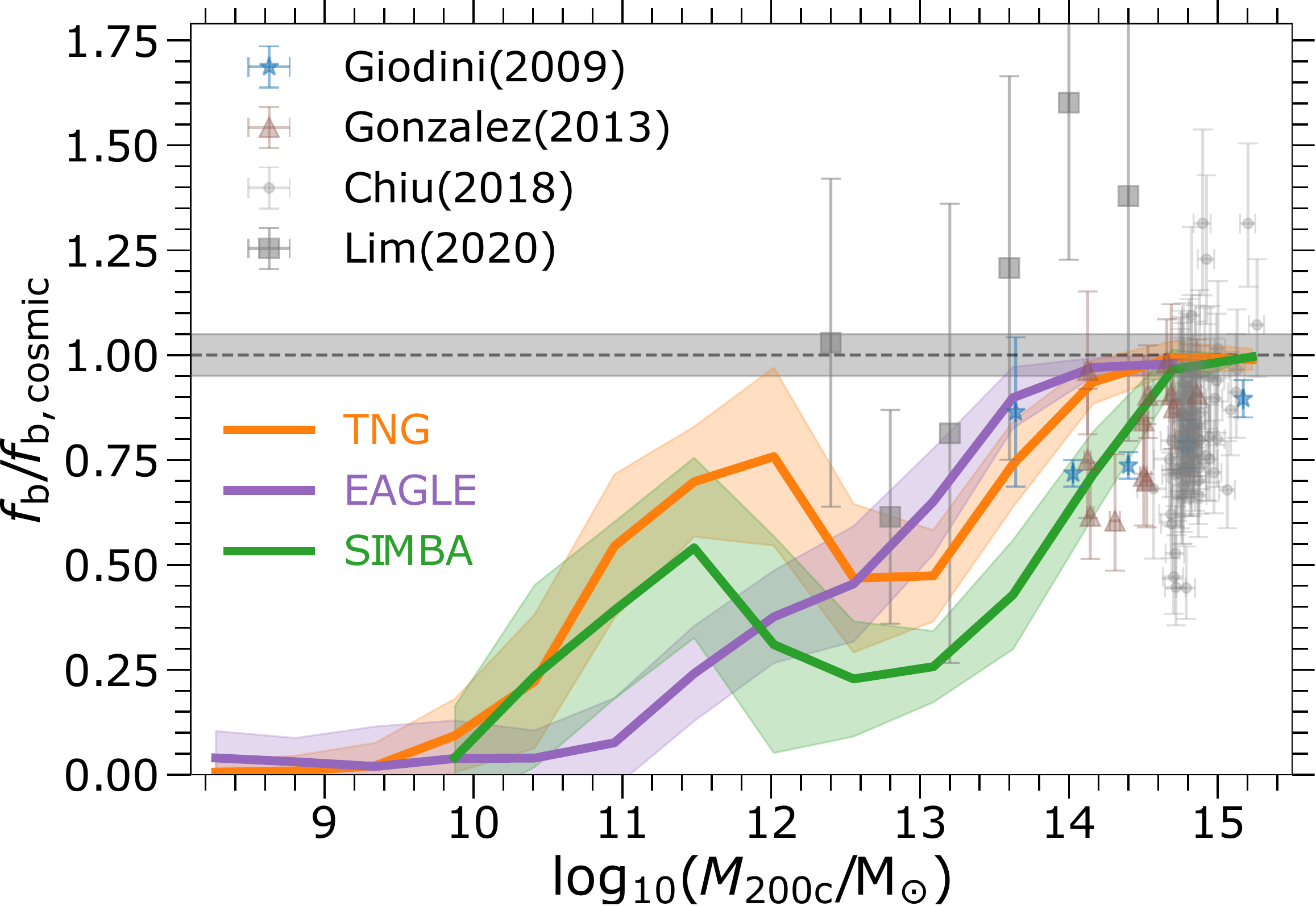}
    \hspace{1mm}
    \includegraphics[width=0.99\columnwidth]{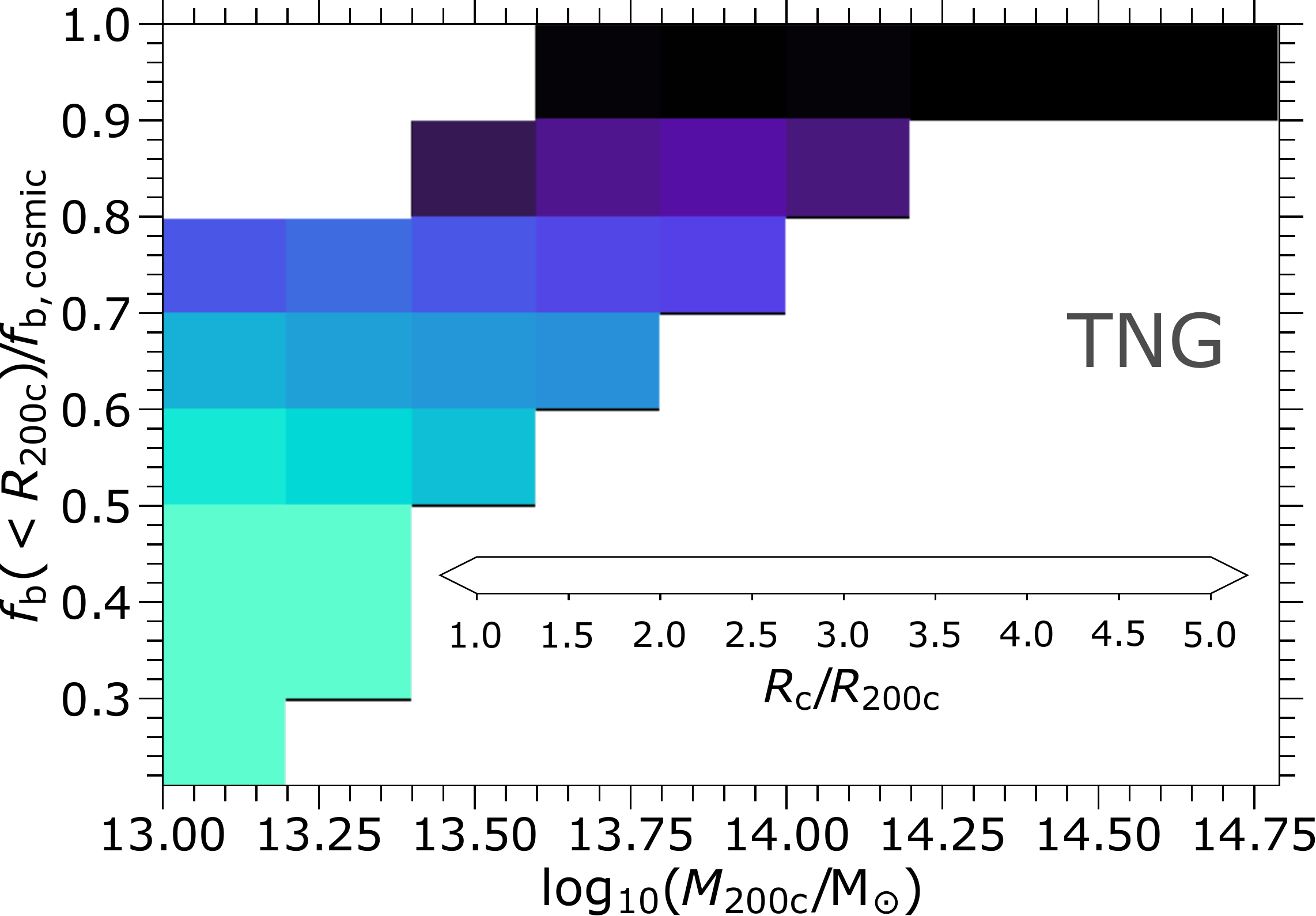}
    \caption{The relations between halo mass, halo baryon fraction and the closure radius at $z=0$. Left panel: halo baryon fraction as a function of halo mass in TNG, EAGLE, and SIMBA, in comparison with X-ray and SZ observations. The X-ray observational data are taken from \protect\cite{giodini2009stellar,gonzalez2013galaxy,chiu2018baryon}, where all correspond to the halo baryon fraction within $R_{\rm 500c}$. The SZ data (squares), a currently challenging observation with large uncertainties, are taken from \protect\cite{lim2020detection} and are reported within $R_{\rm 200c}$. Right panel: 2D histogram of the closure radius $R_{c}/R_{\rm 200c}$ (colours) as a function of the halo mass (y-axis) and halo baryon fraction (x-axis) in TNG. At fixed halo mass (x-axis), there is a strong correlation between the closure radius (colours) and halo baryon fraction (y-axis), which motivates our universal relationship (see text).}
\label{Fig: halo_baryon_fraction}
\end{figure*}

We have previously focused on the average trend of closure radius $R_{\rm c}$ with halo mass. At fixed halo mass, however, this characteristic scale has a non-negligible scatter (see Fig. \ref{Fig: R_c}). We now demonstrate that the dominant secondary parameter, beyond mass, that explains this scatter is the halo baryon fraction. 

Firstly, as previously shown, the halo baryon fraction depends on halo mass, but differently so depending on the simulation model. The left panel of Fig. \ref{Fig: halo_baryon_fraction} shows the halo baryon fraction as a function of halo mass in TNG, EAGLE, and SIMBA compared to available X-ray and SZ observations, for reference. The X-ray data show the baryon fraction within $R_{\rm 500c}$ and are taken from \protect\cite{giodini2009stellar,gonzalez2013galaxy,chiu2018baryon}. The SZ data \citep[squares;][]{lim2020detection} are reported within $R_{\rm 200}$.

We clearly see the trend with the halo mass in all three models, which is caused by the same physical processes that impact the closure radius, i.e. UV background, stellar feedback, and AGN feedback (see Section \ref{subsec: closure_radius}). The simulations agree with each other only for the least and most massive haloes, where feedback processes do not play a critical role in changing the distribution of baryons. The simulations are also broadly consistent with the observed data of galaxy clusters. Nevertheless, substantial discrepancies exist in galaxy groups \citep[see also][]{Oppenheimer2021Simulating}. However, we do not suggest a quantitative comparison between this halo-scale data and the simulations. This would require the careful construction of mock X-ray and SZ observables, which we will undertake in future work \citep[see also][]{truong20,Lim2021Properties}.

We note that the SZ-based observations (squares in Fig. \ref{Fig: halo_baryon_fraction}) infer that haloes have baryon fractions at, or above, the cosmic value, for most haloes with $\log_{10}(M_{\rm 200c}/{M_{\odot}})>12.5$. In some cases ($\log_{10}(M_{\rm 200c}/{\rm M_{\odot}}>13.75$), the halo baryon fraction exceeds the cosmic value significantly, by up to $60\%$. Given the considerable uncertainties and large error bars, we suspect this is due to both uncertainties in the SZ observations, the low angular resolution, their reduction and analysis, as well as the method used to measure the halo mass via the stellar mass to halo mass ratio of the SDSS galaxies. Clearly, the inference of, and comparison with, SZ-based baryon fraction data is challenging, and we postpone careful SZ mock catalogues for future work.

In the right panel of Fig. \ref{Fig: halo_baryon_fraction}, we show median closure radius $R_{\rm c}/R_{\rm 200c}$ (colours) as a function of halo mass (y-axis) and halo baryon fraction (x-axis) for TNG haloes with $\log_{10}(M_{\rm 200c}/M_{\odot})>13$ at $z=0$. At fixed halo mass, $R_{\rm c}/R_{\rm 200c}$ shows a significant trend with the halo baryon fraction, $f_{\rm b}(<R_{\rm 200c})/f_{\rm b, cosmic}$, as $R_{\rm c}/R_{\rm 200c}$ increases with decreasing $f_{\rm b}(<R_{\rm 200c})/f_{\rm b,cosmic}$. On the other hand, at a given halo baryon fraction bin, there is no significant trend with the halo mass. The same argument holds for EAGLE and SIMBA, and when replacing $R_{\rm 200c}$ by $R_{\rm 500c}$. We note that $R_{\rm c}$ itself depends on both the halo mass and halo baryon fraction. However, when normalising the closure radius to the halo virial radius, i.e. $R_{\rm c}/R_{\rm 200c}$, there is no significant residual trend on the halo mass at fixed $f_{\rm b}$. The same finding holds also for higher redshifts.

To estimate the closure radius $R_{\rm c}$ for groups and clusters of $\log_{10}(M_{\rm 200c}/{\rm M_{\odot}})\geq 13$ with a given halo mass, baryon fraction, and redshift, we introduce a simple formula:
\begin{equation}
    \label{eq: universal_r_c}
    \frac{R_{\rm c}}{R_{\rm 200c,500c}} - 1 = \beta(z) \left[1 - \frac{f_{\rm b}(<R_{\rm 200c,500c})}{f_{\rm b,cosmic}}\right],
\end{equation}
where $R_{\rm 200c,500c}$ is the halo $R_{\rm 200c}$ or $R_{\rm 500c}$. Here $\beta(z)$ is a simple power law scaling of redshift:
\begin{equation}
    \label{eq: beta}
    \beta(z) = \alpha\,(1+z)^\gamma.
\end{equation}
We obtain the free parameters $\alpha$ and $\gamma$ by fitting to each simulation. The best fit values are given in Table \ref{tab: universal_r_c}. 

\begin{table}
    \centering
    \caption{The best fit values for the free parameters in the universal relation between the halo baryon fraction and the closure radius, i.e. the characteristic radius within which all baryons associated with the halo are found (Eqs. \ref{eq: universal_r_c},\ref{eq: beta}).}
    \label{tab: universal_r_c}
    \begin{tabular}{|*{3}{c|}}
        \hline \hline
        Model\,/\,Formula & Eqs. \ref{eq: universal_r_c},\ref{eq: beta} with $R_{\rm 200c}$ & Eqs. \ref{eq: universal_r_c},\ref{eq: beta} with $R_{\rm 500c}$ \\
        \hline \hline
        TNG & $\alpha = 8.08, \,\,\, \gamma = -\,0.44$ & $\alpha = 10.69, \,\,\, \gamma = -\,0.33$\\
        \hline
        EAGLE & $\alpha = 7.48,\,\,\, \gamma = -\,0.50$ & $\alpha = 8.86,\,\,\, \gamma = -\,0.31$\\
        \hline
        SIMBA & $\alpha = 14.22,\,\,\, \gamma = -\,0.85$ & $\alpha = 21.31,\,\,\, \gamma = -\,0.84$\\
        \hline
    \end{tabular}
\end{table}
\begin{figure*}
    \centering
    \includegraphics[width=0.49\textwidth]{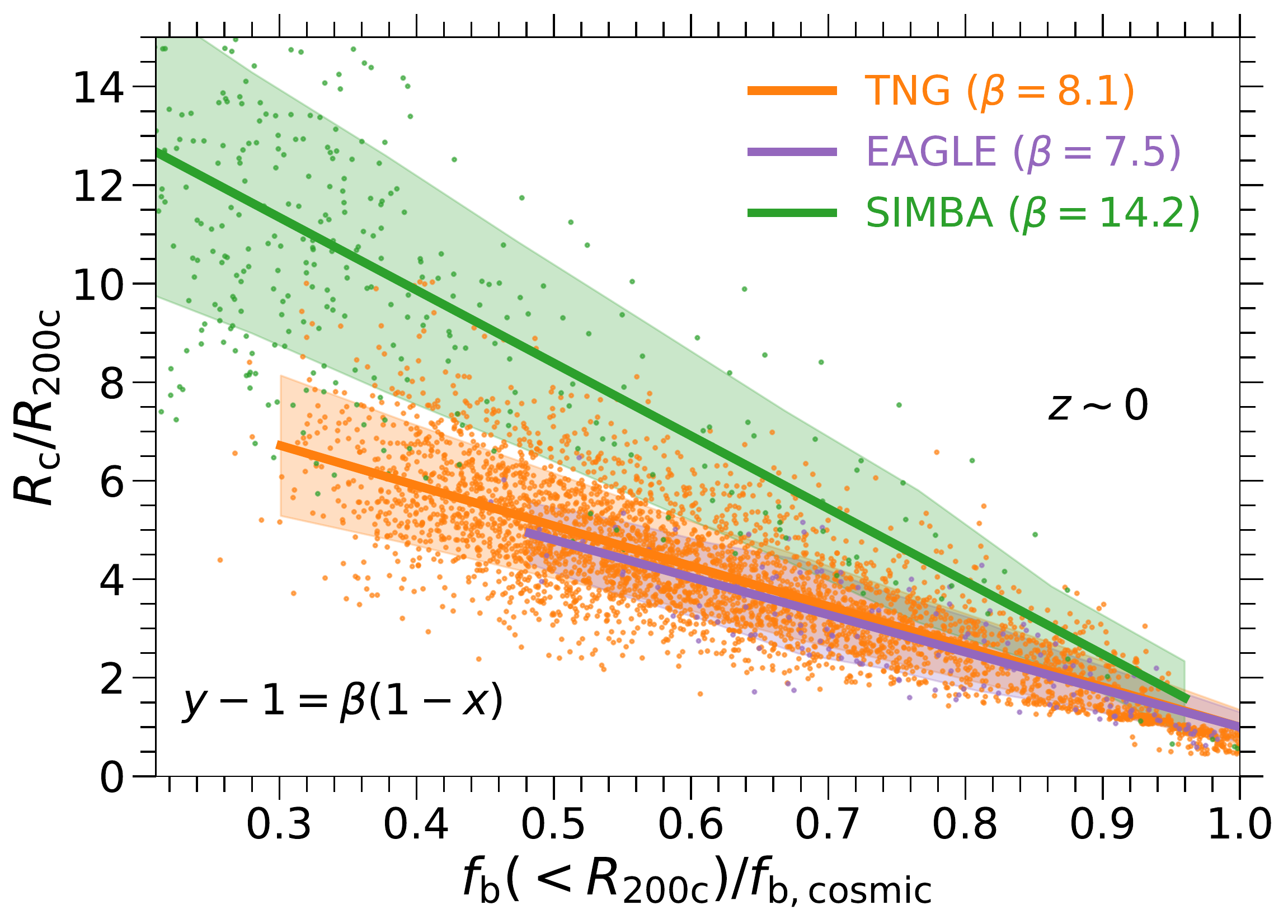}
    \includegraphics[width=0.49\textwidth]{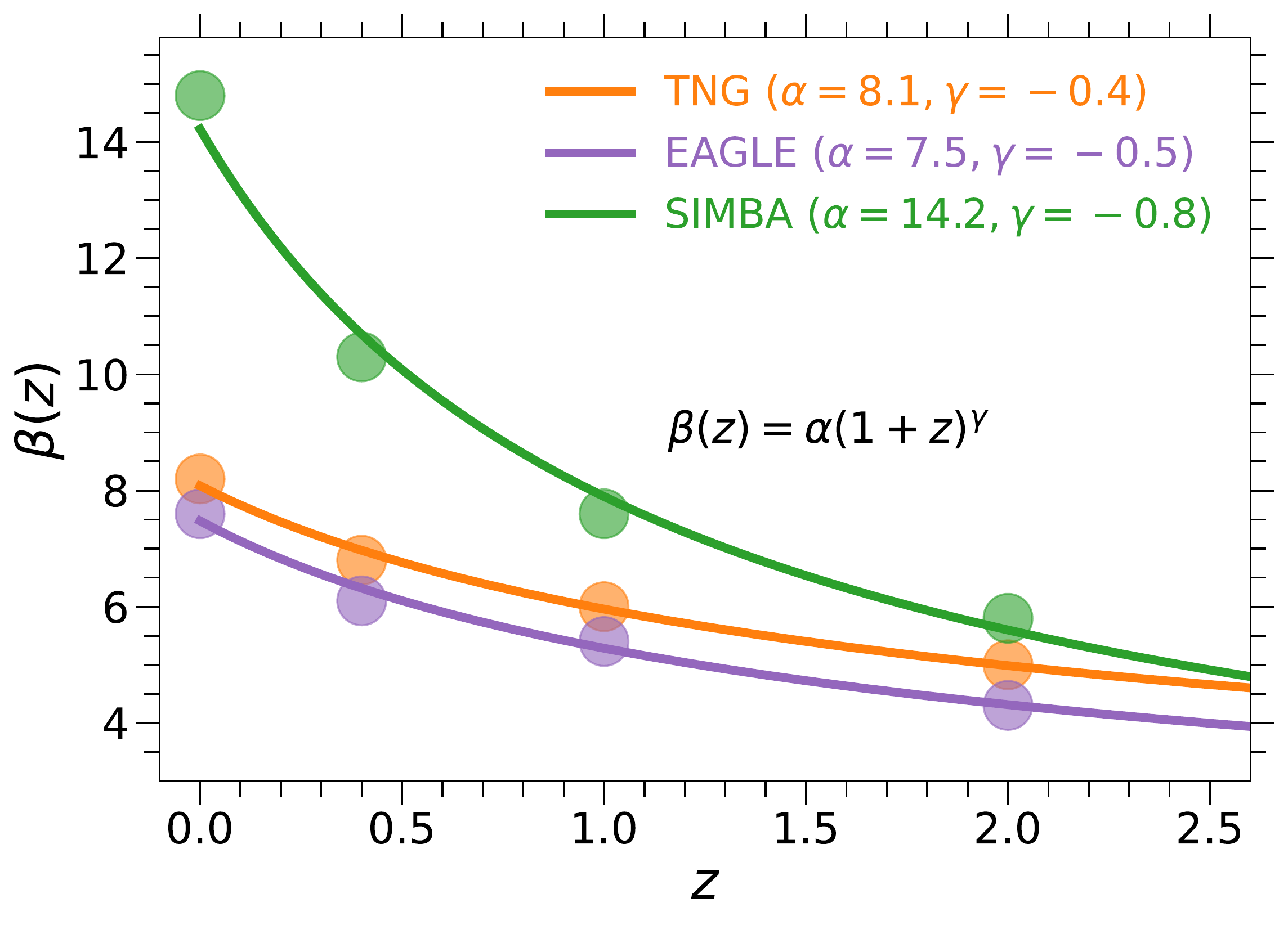}
    \caption{Correlation between the halo baryon fraction and the closure radius $R_{c}$. Left: Normalised closure radius as a function of the halo baryon fraction within $R_{\rm 200c}$ at three different redshifts. Our universal fitting formula, $R_{\rm c}/R_{\rm 200c,500c} - 1 = \beta(z) [1 - f_{\rm b}(<R_{\rm 200c,500c})/f_{\rm b,cosmic}]$, is valid for groups and clusters with $\log_{10}(M_{\rm 200c}/{\rm M_{\odot}})\gtrsim 13$ of all simulations considered: TNG, EAGLE, and SIMBA. The solid lines show the best fit for each simulation, and the shaded regions correspond to the $1\sigma$ of the scatter around the best fit. Replacing $M_{\rm 200c}$ and $R_{\rm 200c}$ with $M_{\rm 500c}$ and $R_{\rm 500c}$ would only alter the slope of our best fit (see Table \ref{tab: universal_r_c}), but leaves the shape of the fitting function and our conclusions unchanged. Right: $\beta(z)$ in our fitting formula (Eqs. \ref{eq: universal_r_c},\ref{eq: beta}) as a function of redshift. The curved lines show our best fit for $\beta(z)$ using the parameters in Table \ref{tab: universal_r_c}. The dots correspond to best values for $\beta(z)$ when fitting Eq. \ref{eq: universal_r_c} at each individual redshift.}
\label{Fig: universal_relation_r_c}
\end{figure*}

In the left panel of Fig. \ref{Fig: universal_relation_r_c}, we show the closure radius $R_{\rm c}/R_{\rm 200c}$ as a function of $f_{\rm b}(<R_{\rm 200c})$ for TNG, EAGLE, and SIMBA at $z=0$. In the right panel, we show how our best fit for $\beta(z)$ (curved lines) compares with the value of $\beta(z)$, fitted at a few individual redshifts ($z=0,0.4,1,2$, dots). Our proposed relation provides a robust fit for all three models, regardless of the significant differences in their physical (feedback) models, as well as numerical methods. Moreover, our fitting relation Eq. \ref{eq: beta} perfectly describes the redshift evolution of the closure radius. The closure radii across the three models become more similar at high redshift, while the scatter in each relation also decreases. We note that estimating $R_{\rm c}/R_{\rm 500c}$ based on $f_{\rm b}(<R_{\rm 500c})$ gives a similarly good fit, although with different slope (see Table \ref{tab: universal_r_c}).

Based on the above, we conclude that the form of Eq. \ref{eq: universal_r_c} is a universal relation between the closure radius $R_c/R_{\rm 200c}$ and the halo baryon fraction for groups and clusters in a $\rm \Lambda CDM$ universe. By ‘universal’ we refer to the unchanged functional form with respect to different hydrodynamical simulations, and with respect to the halo mass. The slope $\alpha$ and redshift dependence $\gamma$ depend on galaxy formation model, suggesting that observations can discriminate between them. Overall, given the value of halo mass (or equally $R_{\rm 200c}$) and halo baryon fraction, one can predict the closure radius $R_{\rm c}$.

Fig. \ref{Fig: universal_relation_r_c} illustrates that different simulations can make substantially different predictions for $R_{\rm c}$. TNG and EAGLE are in fact almost identical at a given halo-scale baryon fraction, while SIMBA clearly ejects baryons to larger distances. The three models we consider produce different values for $R_{\rm c}/R_{\rm 200c}$ as a function of halo mass and halo baryon fraction. Although simulation models may result in similar halo-scale baryon fractions, they can redistribute these baryons, or more specifically gas, to markedly different scales. This causes degeneracy in the modelling of AGN feedback, which however, according to our findings, can be resolved by comparing the models with observations of baryon fraction in the outskirts of haloes.

The redistribution of baryons due to AGN feedback can change the matter power spectrum as well \citep[e.g.][]{springel2018first}. Recently, \cite{vanDaalen2020Exploring} found that at the scale $k<1 \, \rm h \, Mpc^{-1}$ the change in the matter power spectrum due to baryonic matter and galaxy evolution can be expressed as an exponential function of the mean halo baryon fraction (normalised to the cosmic value) in groups and clusters with $\log_{10}M_{200c}/{\rm M_{\odot}}\gtrsim 13$ \citep[see also][]{Schneider2019}. They contrast their model against several cosmological simulations, including TNG and EAGLE. However, they did not consider SIMBA, which shows considerable differences with TNG and EAGLE in the coefficients of our universal relation of the closure radius (Fig. \ref{Fig: universal_relation_r_c}). Given that TNG and EAGLE follow a similar relation both in our study (i.e. similar fits in Fig. \ref{Fig: universal_relation_r_c}) and in \cite{vanDaalen2020Exploring}, whereas SIMBA is an outlier, we suspect that the impact of baryons on $P(k)$ in SIMBA may similarly differ, at fixed group-scale halo gas fraction. Our closure radius, which is driven by large-scale baryon redistribution, is clearly related to the impact of baryonic feedback processes at certain $k$-scales in the matter power spectrum.

\subsection{The closure radius in observations}
\label{subsec: obs}

\begin{figure*}
    \centering
    \includegraphics[width=0.8\textwidth]{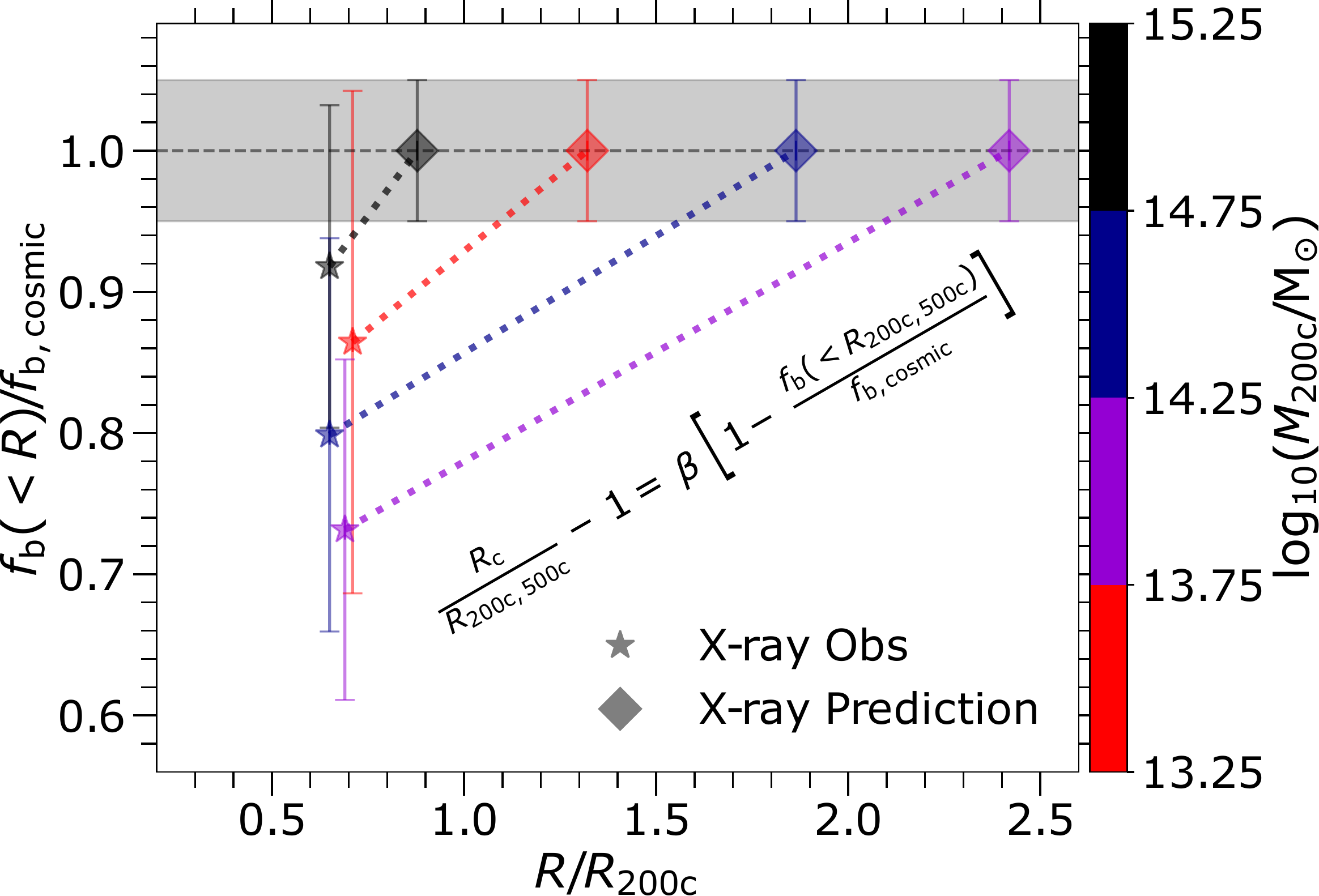}
    \caption{Existing observational estimates of the baryon fraction within haloes and predictions for future observations of average baryon fraction, normalised to the cosmic value, as a function of halocentric distance. The four stars show literature X-ray observations, where we have combined data from the following datasets: \protect\cite{giodini2009stellar,gonzalez2013galaxy,chiu2018baryon}. The large diamonds show our predictions for where future observations will infer $f_{\rm b}=f_{\rm b,cosmic}$, i.e. the distance where baryonic closure is attained. These predictions are based on Eq. \ref{eq: universal_r_c}, and therefore give the expectation for each of these \textit{particular} four combined observational samples, given their mean halo mass and halo $f_{\rm b}$.}
\label{Fig: Profile_f_b_obs}
\end{figure*}

The outskirts of haloes and, in general, the regions beyond the halo boundary are usually poorly observed due to the difficulty of detecting such low-density baryons. Within the halo boundary ($R_{\rm 200c}$ or $R_{\rm 500c}$) of groups and clusters ($\log_{10}M_{\rm 200c}/{M_{\odot}}>13$), however, the baryon fraction is statistically measured in X-ray and SZ observations \citep[e.g. see][]{sanderson2013baryon,chiu2018baryon,lim2020detection}. In this subsection, we use our findings and Eq. \ref{eq: universal_r_c} to make predictions for the closure radius $R_{\rm c}$ of actually observed haloes. 

Fig. \ref{Fig: Profile_f_b_obs} shows the halo baryon fraction as a function of halocentric distance (x-axis) and halo mass (colours). We include X-ray observations (stars), each matched with our prediction for the corresponding closure radius (diamonds). Here, we combine three X-ray observations, including \cite{giodini2009stellar,gonzalez2013galaxy,chiu2018baryon} by averaging over the baryon fraction of haloes within each given halo mass bin.\footnote{Similar to the methods for the galaxy stellar mass function and quenched fractions in \cite{Henriques2013Simulations,Ayromlou2021Galaxy}.} Based on these four samples, each corresponding to a different inferred halo mass and $f_{\rm b}/f_{\rm b,cosmic}$, we then apply the relation previously discussed (Eq. \ref{eq: universal_r_c}) to predict the closure radius $R_{\rm c}/R_{\rm 200c}$. That is, we estimate how far from these \textit{particular} halo samples observations will need to reach in order to find the missing baryons. The resulting predictions, based on the TNG simulations, are shown with large diamonds.
Overall, we predict that the baryons associated with the observed groups and clusters ($\log_{10}M_{\rm 200c} / \rm M_{\odot} > 13$) will be found within $R_{\rm c}<1.5 - 2.5 R_{\rm 200c}$.

These are the expectations from Eq. \ref{eq: universal_r_c} using the TNG simulation best-fit parameter values. The EAGLE model would also predict $R_{\rm c}/R_{\rm 200c}$ in good agreement with TNG. In contrast, SIMBA predicts significantly larger $R_{\rm c}/R_{\rm 200c}$ values, due to its stronger and more ejective AGN feedback.\footnote{Relatedly, in a recent study, \cite{Yang2022SZSimbaTNG} suggest that the jet mode of SIMBA's AGN feedback might be too strong. They investigated the y-decrement, within $R_{\rm 200c}$, versus halo mass (Y-M) relation in TNG and SIMBA and argued that the jet mode of SIMBA's black hole feedback causes substantial deviation from the self-similar expectations of the Y-M relation.} Therefore, observing the baryon fraction in the outskirts of haloes is a promising avenue to break halo-scale degeneracies while constraining AGN feedback models.

\section{Summary and Conclusions}
\label{sec: summary}

The complex distribution of baryons in the Universe leaves important clues about key astrophysical processes. In the framework of $\rm \Lambda CDM$ cosmology, cold dark matter interacts only through gravity, and the distribution of dark matter and the formation of dark matter structures are relatively well understood \citep[][]{navarro1997universal,Springel2005Modelling,Wang2020Universal}. On the other hand, baryonic matter has a more complex thermodynamical evolution, being subject to forces beyond gravity alone. As a result, understanding the distribution of baryons and the physical processes affecting them is a pressing topic in our understanding of galaxy formation and evolution \citep[see][for a comprehensive discussion]{mo2010galaxy}.

In this paper, we employ three sets of cosmological hydrodynamical simulations (TNG50, TNG100 and TNG300 in addition to EAGLE and SIMBA) to study the distribution of baryons and discover the physical processes influencing them both within, and beyond, the boundaries of dark matter haloes. We analyse haloes over a mass range of seven orders of magnitude, spanning the full halo mass range over which the process of galaxy formation occurs ($10\lesssim \log_{10}M_{\rm 200c}/{\rm M_{\odot}}\lesssim 15$), as well as entirely dark low-mass haloes ($8\lesssim \log_{10}M_{\rm 200c}/{\rm M_{\odot}}\lesssim 10$). Moreover, we consider the large-scale environment of each halo, from its centre to scales as large as $30\,R_{\rm 200c}$, i.e. to distances of $\gtrsim 30 \, \rm Mpc$ away from clusters.

We first analyse the cumulative baryon fraction as a function of halocentric distance. Our main results are as follows:

\begin{itemize}
    \item In all three simulations, the cumulative baryon fraction is lower than the cosmic value within the halo boundary, $R_{\rm 200c}$, and increases with halocentric distance at $R\gtrsim R_{\rm 200c}$ (Fig. \ref{Fig: Profile_f_b_c})
    \item There is a characteristic scale within which all baryons associated with haloes are found -- the \textbf{closure radius} $R_{\rm c}$ (Fig. \ref{Fig: R_c}). 
    \item The value of this closure radius depends strongly on halo mass. It also differs among TNG, EAGLE and SIMBA, primarily due to their different physical (i.e. feedback) models (Fig. \ref{Fig: R_c}), and can range from within $R_{\rm 200c}$ (for clusters) out to scales larger than $10\, R_{\rm 200c}$ (for low-mass haloes).
    \item The closure radius is redshift-dependent, and $R_{\rm c}/R_{200c}$ always decreases with increasing redshift, in all the simulations (Fig. \ref{Fig: R_c_z}).
\end{itemize}

Studying variations of the TNG model, we then uncover the dominant physical processes impacting the distribution of baryons and the amplitude of $R_{\rm c}$ at different halo mass ranges. These variants include runs with no stellar feedback, no AGN feedback, no cooling, non-radiative, and combinations therein. We find that:

\begin{itemize}
    \item Galaxy clusters ($\log_{10}M_{\rm 200c}/{M_{\odot}}\gtrsim 14$) are the least influenced by different astrophysical processes, and they typically contain all their baryons within the halo boundary, i.e. $R_{\rm c}\lesssim R_{\rm 200c}$. Their small-scale structure, however, is affected by feedback processes (Fig. \ref{Fig: Profile_f_b_c}).
    \item In galaxy groups ($12 \lesssim \log_{10}M_{\rm 200c}/{M_{\odot}}\lesssim 14$), AGN feedback is the dominant physical process, which ejects the gas beyond the halo boundary, lowering the halo baryon fraction. The baryons missing from these haloes can be found at different halocentric distances in their outskirts (Fig. \ref{Fig: TNG_VAR_R_c}).
    Quantitatively, we find $R_{\rm c}/R_{\rm 200c} \sim 4.5$ (TNG), $\sim 4$ (EAGLE), $\sim 13.5$ (SIMBA) for haloes with $\log_{10} (M_{\rm 200c}/{\rm M_{\odot}})\sim 13$.
    \item Low-mass haloes ($10 \lesssim \log_{10}M_{\rm 200c}/{\rm M_{\odot}}\lesssim 12$) are primarily influenced by stellar feedback, in the form of galactic-scale, and halo-scale, outflows.
    We do not find a noticeable impact from black hole feedback in this halo mass range (Fig. \ref{Fig: TNG_VAR_R_c}).
    For reference, $R_{\rm c}/R_{\rm 200c} \sim 6$ (TNG), $\sim 10$ (EAGLE), $\sim 8.5$ (SIMBA), for haloes with $\log_{10} (M_{\rm 200c}/{\rm M_{\odot}})\sim 11$.
    \item In dark haloes ($8 \lesssim \log_{10}M_{\rm 200c}/{\rm M_{\odot}}\lesssim 10$), which typically do not host a galaxy, UV/X-ray background radiation prevents the gas from accreting onto the halo. In this mass range, $R_{\rm c}/R_{\rm 200c}\gtrsim 10$, on average (Fig. \ref{Fig: TNG_VAR_R_c}).
\end{itemize}

Observations suggest that a considerable fraction of baryons are ``missing'' from dark matter haloes. Our results demonstrate that the baryons missing from simulated haloes reside in their outskirts: namely, within the closure radius. Studying the dependence of this scale on halo and galaxy properties, we find that:

\begin{itemize}
    \item At fixed halo mass for groups and clusters, scatter in the closure radius is explained by a strong correlation with the halo baryon fraction $f_{\rm b}(<R_{\rm 200c})$. Once normalised by the halo radius $R_{\rm 200c}$, the closure radius is independent of halo mass at fixed baryon fraction (Fig. \ref{Fig: halo_baryon_fraction}, right panel).
    \item We introduce a universal relation between the closure radius and halo baryon fraction for groups and clusters: $R_{\rm c}/R_{\rm 200c,500c} - 1 = \beta(z) (1 - f_{\rm b}(<R_{\rm 200c,500c})/f_{\rm b,cosmic})$. Here $\beta(z) = \alpha\,(1+z)^\gamma$, and $\alpha$ and $\gamma$ are free parameters which can be fit using each simulation (Eqs. \ref{eq: universal_r_c},\ref{eq: beta} and Fig. \ref{Fig: universal_relation_r_c}).
    \item We demonstrate how this relationship can be used observationally. Applying it to X-ray inferences of $f_{\rm b}$ in groups and clusters, we predict that all baryons associated with these observed haloes will be found within $R_{\rm c}/R_{\rm 200c} \sim 1.5-2.5$.
    \item In all three models, $R_{\rm c}/R_{\rm 200c}$ is systematically larger in quenched haloes than in star-forming ones, suggesting that $R_{\rm c}$ is influenced by the same processes that reduce star formation in galaxies (Fig. \ref{Fig: R_c_galprop}, left panel).
    \item At fixed halo mass, haloes with undermassive black holes have significantly smaller $R_{\rm c}$ than average, with some haloes having $R_{\rm c}/R_{\rm 200c}\lesssim 1$. This supports the role of ejective AGN feedback as the primary driver for reshaping the baryonic distribution in and around galaxy groups (Fig. \ref{Fig: R_c_galprop}, right panel).
\end{itemize}

On the theoretical side, large-scale high-resolution simulations with reliable physical models are necessary to make robust theoretical predictions for the distribution of baryons in haloes, filaments, and the rest of the cosmic web. As the closure radius $R_{\rm c}$ varies substantially from simulation to simulation, mainly due to different physical feedback models, observations of the baryon distribution out to $R_{\rm c}$ will directly constrain these uncertain physics. In particular, currently-degenerate differences among models will be resolved by observing the outskirts of haloes (see \ref{subsec: obs}). However, this is a challenging regime, and current X-ray and SZ telescopes do not yet routinely detect low-density gas in the outskirts of haloes, much less to even larger scales.
Observing these scales can benefit from data from recent surveys (e.g. eROSITA, \citealt{Predehl2021eROSITA}), and will require future X-ray missions such as Line Emission Mapper X-ray Probe \citep[LEM,][]{Kraft2022} and Athena \citep[][]{Barret2016Athena}, as well as SZ observations like CMB-S4 \citep[][]{Abazajian2019CMB}, CMB-HD \citep[][]{Sehgal2019CMB-HD}, and CORE \citep[][]{Melin2018Dust}.

\section*{Data Availability}

The outputs of the TNG (\href{https://www.tng-project.org/}{https://www.tng-project.org/}), EAGLE (\href{http://icc.dur.ac.uk/Eagle/}{http://icc.dur.ac.uk/Eagle/}), and SIMBA (\href{http://simba.roe.ac.uk/}{http://simba.roe.ac.uk/}) simulations are all publicly available.
We note that herein we have analyzed versions of SIMBA and EAGLE which have been re-processed to enable an apples to apples comparison with TNG, not the original versions (see Section~\ref{sec: Methodology}).
The datasets generated in this paper (e.g., density profiles, baryon fractions, $R_{\rm c}$) will be shared by the corresponding author upon reasonable request.

\section*{Acknowledgements}

MA thanks Julio Navarro, Carlos Frenk, Simon White, Guinevere Kauffmann, Volker Springel, Chris Byrohl, and Robert Yates for fruitful discussions.
MA and DN acknowledge funding from the Deutsche Forschungsgemeinschaft (DFG) through an Emmy Noether Research Group (grant number NE 2441/1-1). This work is supported by the Deutsche Forschungsgemeinschaft (DFG, German Research Foundation) under Germany's Excellence Strategy EXC 2181/1 - 390900948 (the Heidelberg STRUCTURES Excellence Cluster).
The primary TNG simulations were run with compute time granted by the Gauss Centre for Supercomputing (GCS) under Large-Scale Projects GCS-ILLU and GCS-DWAR on the GCS share of the supercomputer Hazel Hen at the High Performance Computing Center Stuttgart (HLRS). Additional simulations and analysis were carried out on the Vera machine of the Max Planck Institute for Astronomy (MPIA) and systems at the Max Planck Computing and Data Facility (MPCDF). 


\vspace{-1em}
\bibliographystyle{mnras}
\bibliography{refbibtex}


\appendix

\section{Resolution convergence}
\label{app: TNG_convergence}

\begin{table}
	\centering
	\caption{The properties of TNG non-radiative runs shown in the bottom panel of Fig. \ref{Fig: app_TNG_resolution_missing_baryons_2dhist}. Here, $n_{\rm NR}$ is the proper number of dark matter particles for the non-radiative run, derived by dividing $M_{\rm 200c} \,(R_{\rm c}=R_{\rm 200c})$ by the mass of each dark matter particle, $m_{\rm DM}$. Note that the TNG100-1-NR resolution simulation is one of our small volume TNG variation boxes. While it has the resolution of TNG100-1, and thus is ideal for this resolution scaling test, it has limited volume, and does not contain massive halos.}
	\label{tab: TNG_resolution_stat}
	\begin{tabular}{|*{4}{c|}}
		\hline \hline
		\textbf{Simulation} & \textbf{$m_{\rm DM}$} & $M_{\rm 200c} \,(R_{\rm c}=R_{\rm 200c})$ & $n_{\rm NR}$ \\
        & $[\rm M_{\odot}]$ & $[\rm M_{\odot}]$ & \\
		\hline \hline
        TNG100-1-NR variant & $7.5\times 10^6$ & $1.2\times 10^{11}$ & $1.6\times 10^4$ \\ \hline
        TNG100-2-NR & $6.0\times 10^7$ & $1.0\times 10^{12}$ & $1.7\times 10^4$ \\ \hline
        TNG100-3-NR & $4.8\times 10^8$ & $7.9\times 10^{12}$ & $1.7\times 10^4$ \\ \hline
		\hline
	\end{tabular}
\end{table}

\begin{figure}
    \centering
    \includegraphics[width=1\columnwidth]{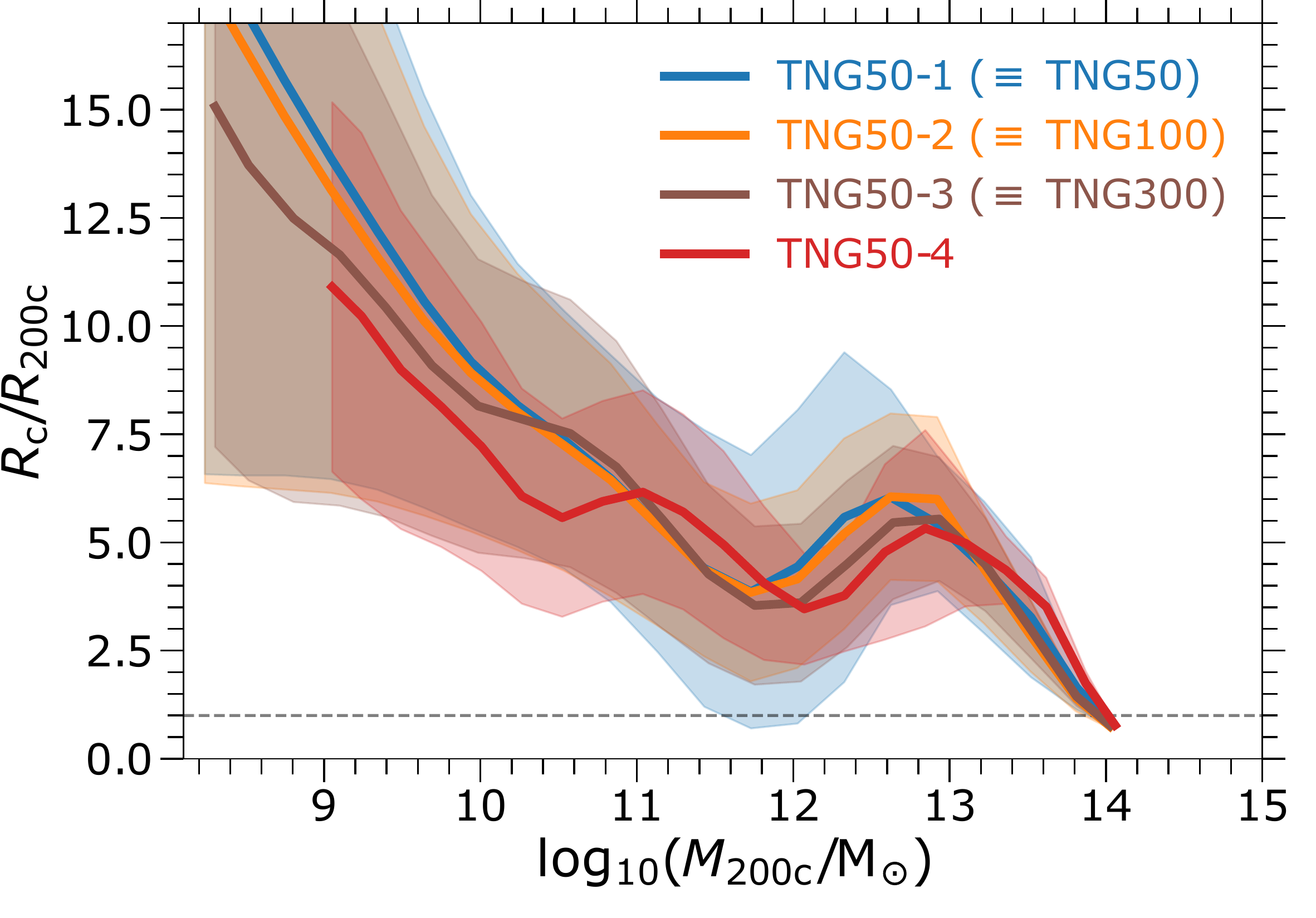}
    \vspace{4mm}
    \includegraphics[width=1\columnwidth]{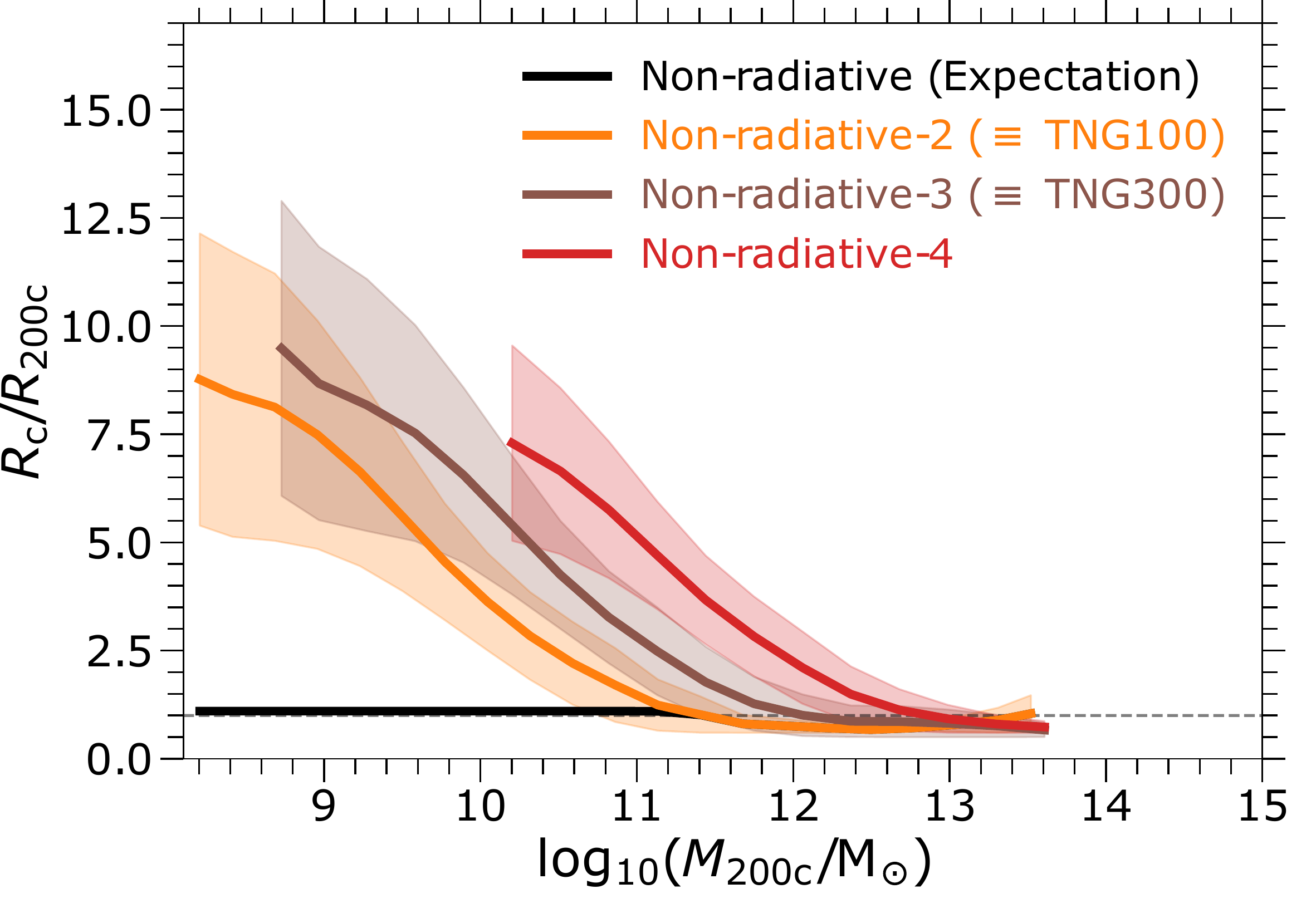}
    \caption{The resolution convergence of our results in TNG. Top: The closure radius in four different runs of the TNG simulation. Each run has a different resolution, with TNG50-1 being the highest resolution and TNG50-2 being the resolution at which the TNG model is calibrated. Bottom: The closure radius in four non-radiative runs of TNG, where each run corresponds to a different resolution. The black line is our expectation of the non-radiative run at a sufficiently high-resolution, i.e. $\gtrsim 10^4$ dark matter particles per halo. The other lines are actual simulations with Non-radiative-2 being the highest resolution one. In both panels, we show the running median value as a function of halo $M_{\rm 200c}$. The shaded regions correspond to 16\% and 84\% percentiles.}
\label{Fig: app_TNG_resolution_missing_baryons_2dhist}
\end{figure}

In Fig. \ref{Fig: app_TNG_resolution_missing_baryons_2dhist} we show $R_{\rm c}/R_{\rm 200c}$ as a function of halo mass for different resolutions of TNG, using the four resolution levels of the TNG50 volume. Each differs by a factor of eight (two) in mass (spatial) resolution. The top and bottom panels show the fiducial and non-radiative models, respectively. In the full physics model (top panel) different resolutions are in excellent agreement with each other. Specifically, our measured values for the closure radius are converging with resolution, and converged to better than $\sim 5-10\%$ between TNG50-2 and TNG50-1 resolutions. In addition, different resolutions overlap at the halo mass boundaries where we switch our analysis from one simulation (resolution) to another. Particularly, we switch from TNG50 (blue) to TNG100 (orange) at $\log_{10}(M_{\rm 200c}/{\rm M_{\odot}})=10$, and switch from TNG100 (orange) to TNG300 (brown) at $\log_{10}(M_{\rm 200c}/{\rm M_{\odot}})=13$.

In the non-radiative simulations (bottom panel), in contrast, resolution convergence at the low-mass end is clearly challenging. In Fig. \ref{Fig: TNG_VAR_R_c}, we used a non-radiative variant with TNG100 resolution but in a smaller box ($l_{\rm box} \sim 37 \, \rm Mpc$, see Table \ref{tab: TNG_resolution_stat}) which has the highest resolution among the TNG non-radiative runs, to date. The solid black line in the bottom panel of Fig. \ref{Fig: app_TNG_resolution_missing_baryons_2dhist} shows our theoretical expectation of a non-radiative run. We remind readers that non-radiative runs are only used to assess the impact of feedback processes, and the slow convergence here is not relevant to our main measurements of the closure radius.

Since there is no UV background to prevent gas from falling onto low-mass haloes, we expect all baryons to be found within the halo boundary. Nevertheless, even our highest resolution (orange line) does not reach this result, presumably due to limited hydrodynamical force resolution. By comparing the mass resolution of each non-radiative run with the halo mass at which each non-radiative simulation starts deviating from $R_{\rm c}/R_{\rm 200c}\leq1$ we estimate the required resolution of a non-radiative run to be $\sim 10^4$ dark matter particles per halo. We summarise this information in Table \ref{tab: TNG_resolution_stat}.

\section{The original Illustris simulation}
\label{app: Illustris}

\begin{figure}
    \centering
    \includegraphics[width=0.95\columnwidth]{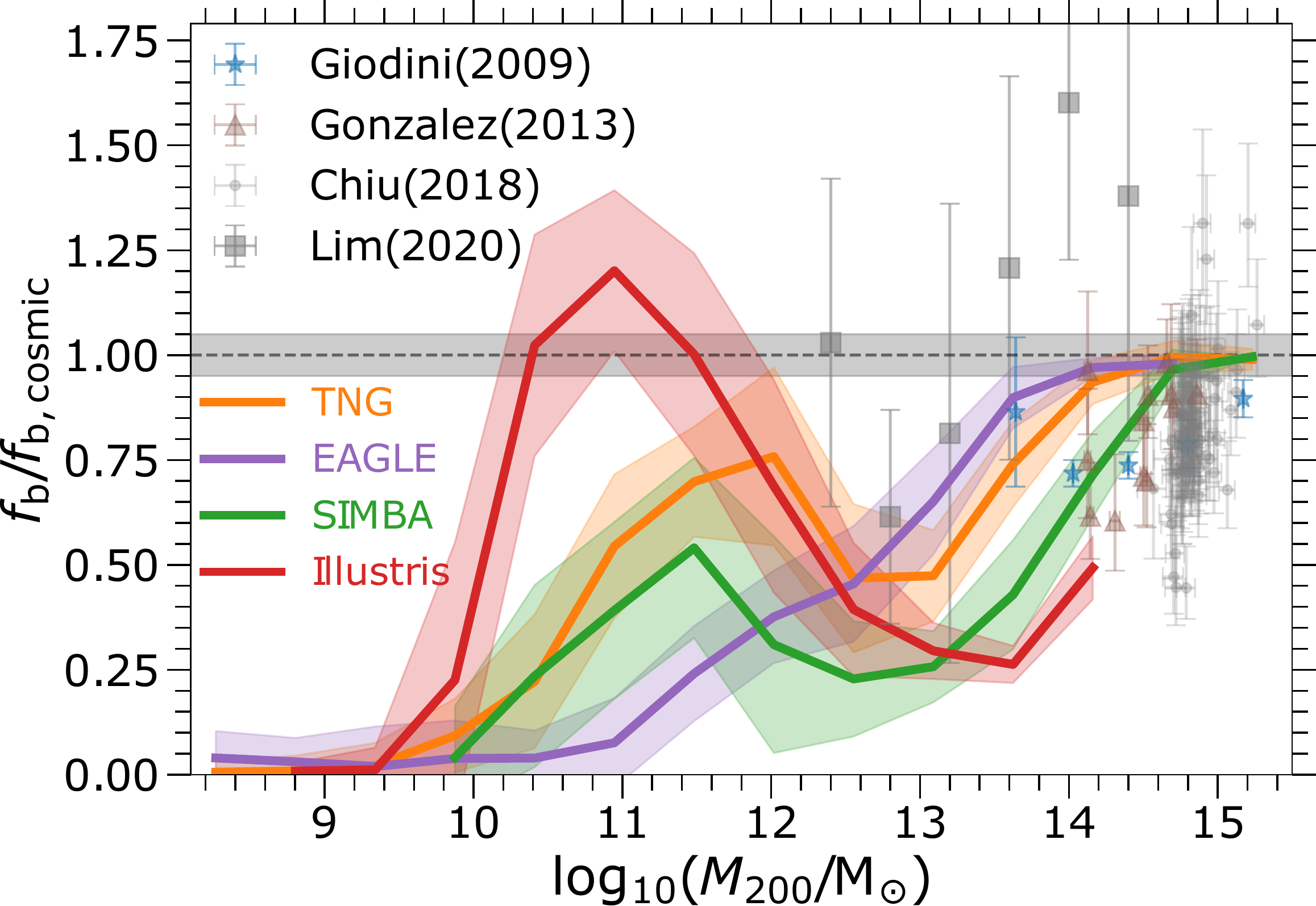}
    \includegraphics[width=0.95\columnwidth]{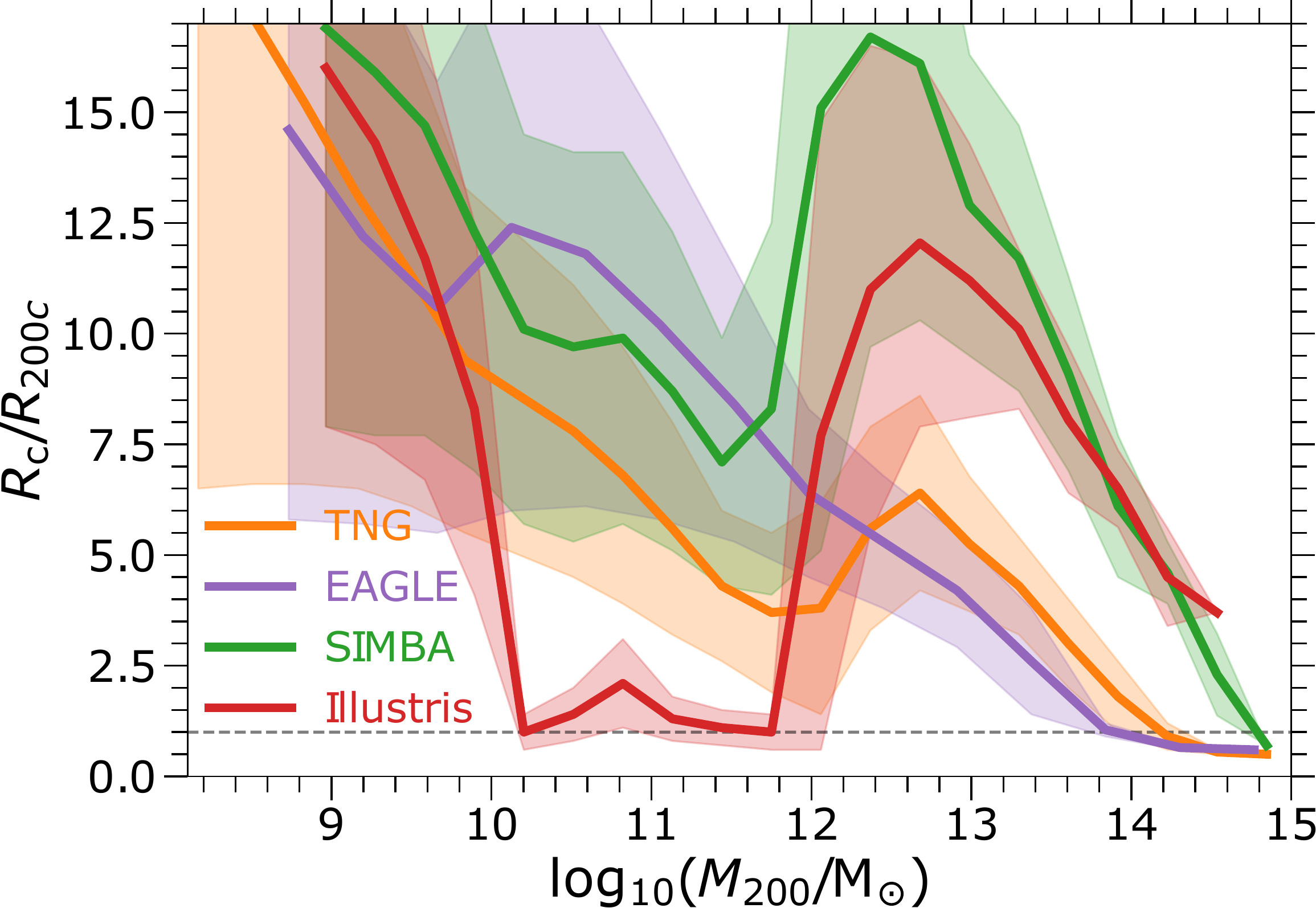}
    \includegraphics[width=0.95\columnwidth]{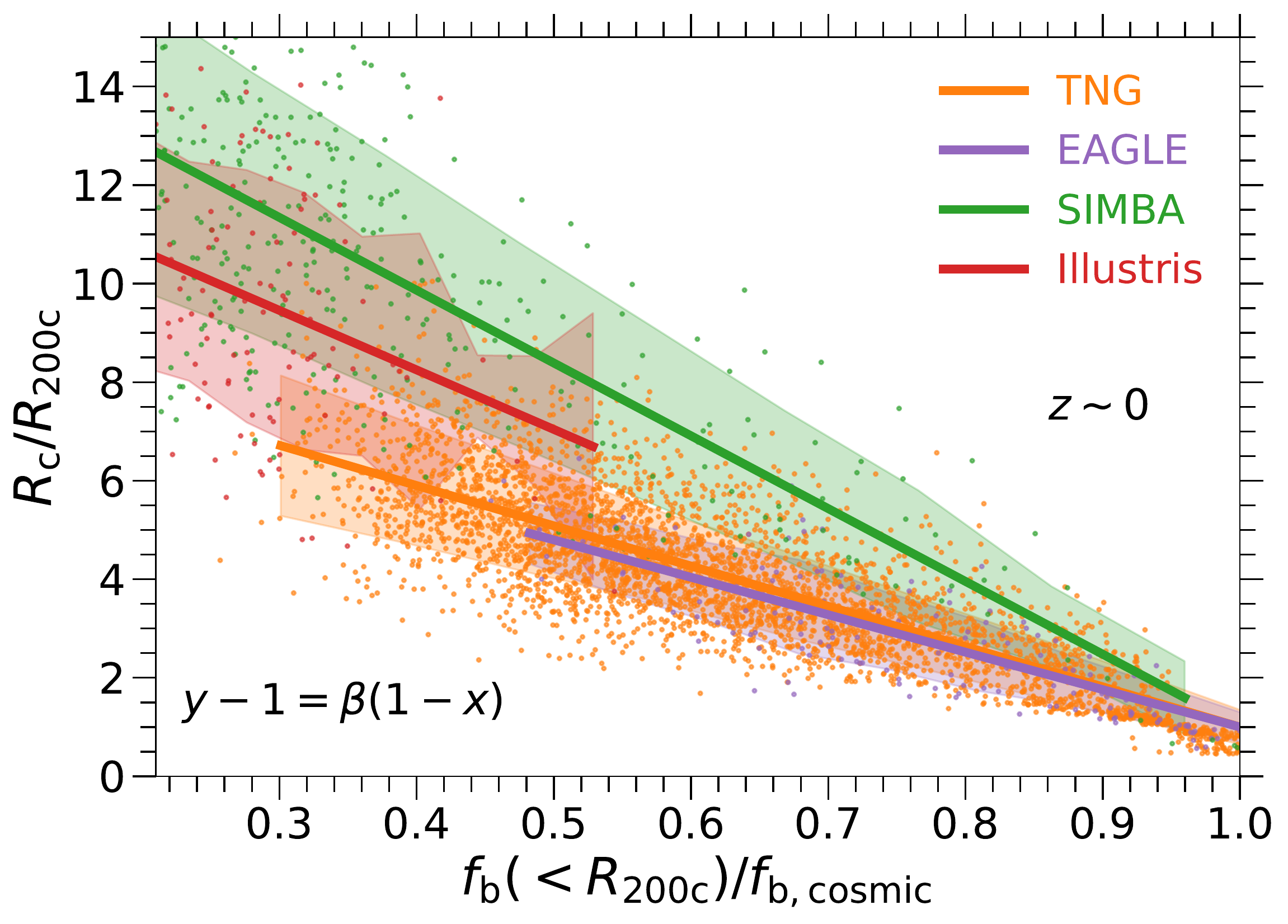}
    \caption{The same analysis as in Figures. \ref{Fig: halo_baryon_fraction}, \ref{Fig: R_c}, \ref{Fig: universal_relation_r_c}, with the original Illustris simulation included. Top: Baryon fraction within $R_{\rm 200c}$ as a function of halo mass. Middle: The closure radius as a function of halo mass. Bottom: Our best-fit relationship for the closure radius as a function of the halo baryon fraction in groups and clusters.}
    \label{Fig: app_Illustris}
\end{figure}

This appendix discusses the key results of this paper, for the original Illustris simulation, which we have omitted from the main text for clarity. The top and middle panels of Fig. \ref{Fig: app_Illustris} show the halo baryon fraction and closure radius as a function of halo $M_{\rm 200c}$. These are analogues to the left panel of Fig. \ref{Fig: halo_baryon_fraction} and the top panel of Fig. \ref{Fig: R_c}. At low halo masses $\log_{10}(M_{\rm 200c}/{\rm M_{\odot}})$, the original Illustris simulation is in reasonable agreement with the three other simulations (TNG, EAGLE, SIMBA) both in halo baryon fraction and in the amplitude of the closure radius.

On the other hand, Illustris haloes with $\log_{10}(M_{\rm 200c}/{\rm M_{\odot}})\sim 11$ have a significantly higher baryon fraction, even slightly exceeding the cosmic value. As a result, one needs to go beyond $R_{\rm 200c}$ in order to reach the cosmic baryon fraction. Consequently, the closure radius here is small, and our method fixes it near unity. There is also small rise in the closure radius at $\log_{10}(M_{\rm 200c}/{\rm M_{\odot}})\sim 11$, which is due to the peak of the halo baryon fraction at the same halo mass.

Overall, this behaviour is starkly different than what we see in the other three simulations, indicating that comparing to TNG, EAGLE, and SIMBA, galactic winds of the original Illustris simulation are less efficient in removing the gas from haloes \citep[see][]{pillepich2018Simulating}. As TNG is in good agreement with low redshift data on the cold gas contents of galaxies, i.e. neutral HI \citep{stevens19,stevens21,diemer19}, this suggests that the original Illustris simulation produces galaxies which are much too gas rich at these mass scales.

In haloes more massive than the Milky Way ($\log_{10}M_{\rm 200c}/{\rm M_{\odot}}\gtrsim 12$), the baryon fraction of Illustris is smaller than TNG and EAGLE and is closer to SIMBA. In this mass range, Illustris predicts a larger closure radius than TNG and EAGLE but smaller than SIMBA, making it an interesting intermediate case. This reflects the different implementations of AGN feedback in these simulations. In particular, the original Illustris simulation implemented a `bubble-mode' feedback at low accretion rates, designed to capture large, jet-inflated bubbles in the hot intracluster medium \citep{sijacki2015illustris}. This mode was found to be overly ejective, in tension with available group-scale halo gas fraction constraints \citep{genel2014introducing}, while at the same time unable to sufficiently quench the galaxy population and produce a reasonable intermediate-mass galaxy optical color bimodality in comparison to SDSS \citep{nelson18a}. We note, however, that a updated parameterization of this underlying model has been used to more success in the FABLE simulations \citep{henden18,henden20}, and it would be useful to compare predictions for the closure radius from FABLE in the future.

Finally, in the bottom panel of Fig. \ref{Fig: app_Illustris}, we show our universal relation (Eq. \ref{eq: universal_r_c}) between the closure radius and halo baryon fraction at $z=0$ (analogues to the right panel of Fig. \ref{Fig: universal_relation_r_c}). Like TNG, EAGLE, and SIMBA, Illustris follows our universal relation at all redshifts. The best-fit parameter values for Eqs. \ref{eq: universal_r_c},\ref{eq: beta} ($\alpha = 11.8, \gamma = -1$ for $R_{\rm 200c}$, and $\alpha = 19.3, \gamma = -1$ for $R_{\rm 500c}$) make it an intermediate case.

\label{lastpage}

\end{document}